\documentclass{article}
\usepackage[dvips]{graphicx}  

\title{{\bf QUANTUM FOAM,  GRAVITY  
 AND GRAVITATIONAL WAVES}}

\author{ {\bf Reginald T. Cahill}\\
School of Chemistry, Physics and Earth Sciences\\
 Flinders University \\
 GPO Box 2100, Adelaide 5001, Australia \\
 Reg.Cahill@flinders.edu.au\\  \\  \\To be published in \\{\bf Relativity, Gravitation,
Cosmology}}

\begin{document}

\date{}

\maketitle

\begin{center} Abstract  \end{center}

{The new information-theoretic Process Physics has shown that space is a quantum foam system
with gravity being, in effect, an inhomogeneous in-flow of the quantum foam into matter.  The theory predicts that
absolute motion with respect to this system should be observable, and   it is shown here  that
 absolute motion has been detected  in at least seven experiments.  As well this experimental data also reveals
the existence of a gravitational wave phenomena associated with the in-flow.  It is shown that Galilean
Relativity and  Special Relativity are in fact compatible, contrary to current beliefs: absolute motion actually
causes the special relativity effects.  The new theory of gravity passes all the tests of the previous Newtonian
and General Relativity theories, but in addition resolves the numerous gravitational anomalies such as the
spiral galaxy `dark matter' effect, the absence of `dark matter' in elliptical galaxies, the inconsistencies in
measuring $G$, the borehole $g$ anomaly, and others.  It is shown that  Newtonian gravity is deeply flawed,
because the solar system from which it was developed has too high a spherical symmetry to have revealed key
aspects of the phenomena of gravity, and that General Relativity inherited these flaws. The data are revealing
that space has structure, and so indicates for the first time evidence of quantum space and quantum gravity
effects.
   }

\vskip12pt 


\newpage

 \tableofcontents

\vskip12pt
\section{  Introduction\label{section:introduction}}
\vskip6pt

The new information-theoretic {\it Process Physics}
\cite{URL,RC01,GQF,AMGE,CK98,CK99,CKK00,CK,MC}  provides for the first
time an explanation of space as a quantum foam system in which gravity is an inhomogeneous flow of the
quantum foam into matter.  That work has implied that absolute motion should be observable and that 
gravity is caused by an effective inhomogeneous in-flow of quantum-foam/space into matter. 
  It is shown here  that  Newtonian gravity and General Relativity   may be re-written in a
`fluid in-flow' formalism, and that a simple generalisation of this  formalism leads to a new theory of
gravity, at the classical level,  that is in better agreement with the experimental data.  It passes all the standard
tests of both the Newtonian and the General Relativity  theories of gravity. Significantly this new theory of gravity
is shown to  resolve the many gravitational anomalies that have been reported -  the spiral galaxy `dark matter'
effect, the absence of `dark matter' in elliptical galaxies, the ongoing inconsistencies in measuring $G$, the
borehole $g$ anomalies and many others. These anomalies, as it is now becoming clear, were revealing  deep flaws in the
Newtonian and General Relativity formalisms.  It turns out that Newtonian gravity is flawed because in its initial
formulation the phenomena of the solar system were too special - the solar system has too much spherical symmetry to
have revealed all the aspects of gravity.  General Relativity, in turn, is seen to be also flawed, because it
`inherited'  these flaws from the Newtonian theory. As well the new theory of gravity predicts a new kind a
gravitational wave, essentially turbulence in the in-flowing space, and this phenomena is evident in the experimental
data.  As well it is shown that Galilean Relativity and the Lorentzian Relativity are actually consistent, and together
describe real physical phenomena - until now they were regarded as mutually exclusive.  Overall we see that the
quantum foam system that is space is more complex and subtle than the models and paradigms of current physics. 
These developments indicate that we  are seeing for the first time evidence of quantum space and quantum gravity
effects - the experimental data is revealing that space has  `structure'.

An analysis  herein  of data from seven experiments
 reveals that absolute motion  relative to space  has been observed  by Michelson and Morley (1887)
\cite{MM}, Miller (1925/26)
\cite{Miller2}, Illingworth (1927) \cite{Illingworth}, Joos (1930) \cite{Joos}, Jaseja {\it et al.}
(1963) \cite{Jaseja}, Torr and Kolen
(1981) \cite{Torr}, and by  DeWitte (1991) \cite{DeWitte}, contrary to common belief  within physics that absolute
motion has never been observed.   The first five of these were Michelson interferometer experiments operating with a
gas, while the last two were coaxial cable RF travel-time experiments using atomic clocks.  Amazingly no-one
had ever analysed the fringe shift data from the interferometer experiments using  two well-known but overlooked key
effects; namely  the Fitzgerald-Lorentz contraction effect and the refractive index effect which  slows
down the speed of light in the gas. The  Dayton Miller  data also reveals the in-flow of space into the
sun which manifests as  gravity.    The experimental data of Miller, DeWitte, and Torr and Kolen indicate that the
in-flow manifests turbulence, which amounts to the observation of a gravitational wave phenomena.

Absolute motion is consistent with special relativistic effects, which are
caused by actual dynamical effects of absolute motion through the quantum foam.  The Lorentzian 
interpretation of relativistic effects  is seen to be essentially  correct.  Vacuum Michelson
interferometer experiments or its equivalent
\cite{KT, BH, Muller, NewVacuum} cannot detect absolute motion.  The various gas-mode Michelson
interferometer data cannot be analysed unless the special relativistic effects are taken into account, and indeed
these experiments demonstrate the validity and reality of the Fitzgerald-Lorentz contraction effect.

\vskip12pt
\section{ A New Theory of Gravity  \label{section:gravity}}
\vskip6pt

\subsection{ Classical Effects of Quantum Foam In-Flow \label{subsection:classicalquantum}}

We begin here the analysis that reveals the new theory and explanation of gravity.  In this theory
gravitational effects are caused solely by an inhomogeneous `flow' of the quantum foam. The new
information-theoretic concepts underlying this physics were discussed in \cite{URL, RC01, CK98}. 
Essentially matter effectively acts as a `sink' for that quantum foam.  It  is important to realise that this is
not a flow of `something' through space; rather it is  ongoing structural changes in space - a fluctuating and
classicalising quantum foam, but with those changes most easily described as a `flow', though such a flow is only
evident from distributed observers.  The Newtonian theory of gravity was based on observations of planetary motion
within the solar system.  It turns out that the solar system was too special, as the planets acted as test objects
in orbit about a spherically symmetric matter distribution - the sun. As soon as we depart from such spherical
symmetry, and even within a spherically symmetric matter distribution problems appear.  Only the numerous, so-far
unexplained, gravitational anomalies are actually providing clues as the the real nature of gravity. The Newtonian
theory was originally formulated in terms of a force field, the gravitational acceleration
${\bf g}({\bf r},t)$,  but as will be shown here it is much closer to the truth if we re-formulate it as a
`fluid-flow' system. The gravitational acceleration
${\bf g}$ in the Newtonian theory is determined by  the matter density
$\rho({\bf r},t)$ according to
\begin{equation}\label{eqn:g1}
\nabla.{\bf g}=-4\pi G\rho.
\end{equation}
For $\nabla \times {\bf g}=0$ this gravitational acceleration ${\bf g}$ may be written as the
gradient of the gravitational potential $\Phi({\bf r},t)$ 
\begin{equation}{\bf g}=-{\bf \nabla}\Phi,\end{equation}  
where the  gravitational
potential is now determined  by $ \nabla^2\Phi=4\pi G\rho $.  Here, as usual, $G$ is the gravitational constant. 
Now as $\rho\geq 0$ we can choose to have 
$\Phi
\leq 0$ everywhere if $\Phi \rightarrow 0$ at infinity. So we can introduce  ${\bf v}^2=-2\Phi \geq 0$ where 
${\bf v}({\bf r},t)$ is some velocity vector field.
  Here the value of ${\bf v}^2$ is
specified, but not the direction of ${\bf v}$. Then
\begin{equation}
{\bf g}=\frac{1}{2}{\bf \nabla}({\bf v}^2)=({\bf v}.{\bf \nabla}){\bf v}+
{\bf v}\times({\bf \nabla}\times{\bf v}).
\label{eqn:f1}
\end{equation} 
For irrotational flow 
${\bf \nabla} \times {\bf v}={\bf 0}$. Then ${\bf g}$ is
the usual Euler expression for the  acceleration of a fluid element in a
time-independent or stationary fluid flow.   If the flow is time dependent and irrotational that expression
is expected to become 
 \begin{equation}{\bf g}=({\bf v}.{\bf \nabla}){\bf v}
+\displaystyle{\frac{\partial {\bf v}}{\partial t}}.
\label{eqn:f2}\end{equation}
Then to be consistent with (\ref{eqn:g1}) in the case of a time-dependent matter density the `fluid flow' form
of Newtonian gravity is 
\begin{equation}
\frac{\partial}{\partial t}(\nabla.{\bf v})+\frac{1}{2}\nabla^2({\bf v}^2)=-4\pi G\rho.
\label{eqn:f3}\end{equation}
This `fluid flow' system has wave-like solutions, in general, but these waves do not manifest as a force via
${\bf g}$. But, as we shall see later, the flow velocity field ${\bf v}$ is observable, and the experimental
data reveals not only ${\bf v}$ but this wave phenomenon.  In the generalisation of (\ref{eqn:f3}), namely 
(\ref{eqn:newgravity}), the wave phenomenon does affect ${\bf g}$.  There is experimental evidence that this
effect has also been observed, as discussed in sect.\ref{subsection:gravitationalanomalies}.  Of course within
the fluid flow interpretation (\ref{eqn:f2}) and (\ref{eqn:f3}) are together equivalent to the Universal Inverse
Square Law for Gravity. Indeed for a spherically symmetric distribution of matter of total mass $M$ the
stationary velocity field outside of the matter 
\begin{equation}
{\bf v}({\bf r})=-\sqrt{\frac{2GM}{r}}\hat{\bf r},
\label{eqn:vfield}\end{equation}
satisfies (\ref{eqn:f3}) and reproduces the inverse square law form for ${\bf g}$ using (\ref{eqn:f2}): 
\begin{equation}
{\bf g}=-\frac{GM}{r^2}\hat{\bf r}.
\label{eqn:InverseSqLaw}\end{equation} 
The in-flow direction
$-\hat{\bf r}$  in (\ref{eqn:vfield}) may be replaced by any other direction, in which case however
the direction of ${\bf g}$ in (\ref{eqn:InverseSqLaw})  remains radial. 

As we shall see of the many new effects
predicted by the generalisation of (\ref{eqn:f3}) one is that this `Inverse Square Law'  is only
valid outside of spherically symmetric matter systems.  Then, for example, the `Inverse Square Law'
is expected to be inapplicable to spiral galaxies. The incorrect assumption of the universal validity
of this law led to the notion of  `dark matter' in order to reconcile the faster observed rotation 
velocities of matter within  such galaxies compared to that predicted by the above law.

To arrive at the new in-flow theory of gravity we require that the velocity field  ${\bf v}({\bf r},t)$ be
specified and measurable with respect to a suitable frame of reference.  We shall use the Cosmic Microwave
Background (CMB) frame of reference for that purpose \cite{CMB}.  Then a `test object' has velocity  
 ${\bf v}_0(t)=d{\bf r}_0(t)/dt$ 
with respect to that CMB frame, where ${\bf r}_0(t)$  is the position of the object
wrt that frame.   We then define 
\begin{equation}
{\bf v}_R(t) ={\bf v}_0(t) - {\bf v}({\bf r}_0(t),t),
\label{eqn:18}\end{equation}
as the velocity of the test object relative to the quantum foam at the location of the object.

Process Physics \cite{URL}  leads to the Lorentzian interpretation of so called `relativistic
effects'.  This means that the speed of light is only `c' with respect to the quantum-foam system, and that time
dilation effects for clocks and length contraction effects for rods are caused by the motion of clocks and rods
relative to the quantum foam. So these effects are real dynamical effects caused by the quantum foam, and are
not to be interpreted as spacetime effects as suggested by Einstein.  To arrive at the dynamical description of
the various effects of the quantum foam we shall introduce conjectures that essentially lead to a
phenomenological description of these effects. In the future we expect to be able to derive this dynamics
directly from the Quantum Homotopic Field Theory formalism \cite{RC01} that emerges from the
information-theoretic system.

First we shall conjecture that the path of an object through an inhomogeneous and time-varying quantum-foam is
determined by a variational principle, namely the path ${\bf r}_0(t)$ minimises the travel time  
\begin{equation}
\tau[{\bf r}_0]=\int dt \left(1-\frac{{\bf v}_R^2}{c^2}\right)^{1/2},
\label{eqn:f4}
\end{equation}  
with ${\bf v}_R$ given by (\ref{eqn:18}). Under a deformation of
the trajectory  ${\bf r}_0(t) \rightarrow  {\bf r}_0(t) +\delta{\bf r}_0(t)$,
${\bf v}_0(t) \rightarrow  {\bf v}_0(t) +\displaystyle\frac{d\delta{\bf r}_0(t)}{dt}$,  and we also
have
\begin{equation}\label{eqn:G2}
{\bf v}({\bf r}_0(t)+\delta{\bf r}_0(t),t) ={\bf v}({\bf r}_0(t),t)+(\delta{\bf
r}_0(t).{\bf \nabla}) {\bf v}({\bf r}_0(t))+... 
\end{equation}
Then
\begin{eqnarray}\label{eqn:G3}
& &\delta\tau=\tau[{\bf r}_0+\delta{\bf r}_0]-\tau[{\bf r}_0]  \nonumber\\
&=&-\int dt \:\frac{1}{c^2}{\bf v}_R. \delta{\bf v}_R\left(1-\displaystyle{\frac{{\bf
v}_R^2}{c^2}}\right)^{-1/2}+...\nonumber\\
&=&\int dt\frac{1}{c^2}\left({\bf
v}_R.(\delta{\bf r}_0.{\bf \nabla}){\bf v}-{\bf v}_R.\frac{d(\delta{\bf
r}_0)}{dt}\right)\left(1-\displaystyle{\frac{{\bf v}_R^2}{c^2}}\right)^{-1/2}+...\nonumber\\ 
&=&\int dt \frac{1}{c^2}\left(\frac{{\bf v}_R.(\delta{\bf r}_0.{\bf \nabla}){\bf v}}{ 
\sqrt{1-\displaystyle{\frac{{\bf
v}_R^2}{c^2}}}}  +\delta{\bf r}_0.\frac{d}{dt} 
\frac{{\bf v}_R}{\sqrt{1-\displaystyle{\frac{{\bf
v}_R^2}{c^2}}}}\right)+...\nonumber
\end{eqnarray}
\begin{equation}
=\int dt\: \frac{1}{c^2}\delta{\bf r}_0\:.\left(\frac{({\bf v}_R.{\bf \nabla}){\bf v}+{\bf v}_R\times({\bf
\nabla}\times{\bf v})}{ 
\sqrt{1-\displaystyle{\frac{{\bf
v}_R^2}{c^2}}}}  +\frac{d}{dt} 
\frac{{\bf v}_R}{\sqrt{1-\displaystyle{\frac{{\bf
v}_R^2}{c^2}}}}\right)+...
\end{equation}
Hence a 
trajectory ${\bf r}_0(t)$ determined by $\delta \tau=0$ to $O(\delta{\bf r}_0(t)^2)$ satisfies 
\begin{equation}\label{eqn:G4}
\frac{d}{dt} 
\frac{{\bf v}_R}{\sqrt{1-\displaystyle{\frac{{\bf v}_R^2}{c^2}}}}=-\frac{({\bf
v}_R.{\bf \nabla}){\bf v}+{\bf v}_R\times({\bf
\nabla}\times{\bf v})}{ 
\sqrt{1-\displaystyle{\frac{{\bf v}_R^2}{c^2}}}}.
\end{equation}
Let us now write this in a more explicit form.  This will
also allow the low speed limit to be identified.   Substituting ${\bf
v}_R(t)={\bf v}_0(t)-{\bf v}({\bf r}_0(t),t)$ and using 
\begin{equation}\label{eqn:G5}
\frac{d{\bf v}({\bf r}_0(t),t)}{dt}=({\bf v}_0.{\bf \nabla}){\bf
v}+\frac{\partial {\bf v}}{\partial t},
\end{equation}
we obtain
\begin{equation}\label{eqn:G6}
\frac{d}{dt} 
\frac{{\bf v}_0}{\sqrt{1-\displaystyle{\frac{{\bf v}_R^2}{c^2}}}}={\bf v}
\frac{d}{dt}\frac{1}{\sqrt{1-\displaystyle{\frac{{\bf v}_R^2}{c^2}}}}+\frac{({\bf v}.{\bf
\nabla}){\bf v}-{\bf v}_R\times({\bf
\nabla}\times{\bf v})+\displaystyle{\frac{\partial {\bf v}}{\partial t}}}{ 
\displaystyle{\sqrt{1-\frac{{\bf v}_R^2}{c^2}}}}.
\end{equation}
Then in the low speed limit  $v_R \ll c $   we  obtain
\begin{equation}{\label{eqn:G7}}
\frac{d{\bf v}_0}{dt}=({\bf v}.{\bf
\nabla}){\bf v}-{\bf v}_R\times({\bf \nabla}\times{\bf v})+\frac{\partial {\bf v}}{\partial t}={\bf g}({\bf
r}_0(t),t)+({\bf \nabla}\times{\bf v})\times{\bf v}_0,
\end{equation}
which agrees with the  `Newtonian' form (\ref{eqn:f2}) for zero vorticity (${\bf \nabla}\times{\bf
v}=0$).   Hence (\ref{eqn:G6}) is a generalisation of (\ref{eqn:f2}) to include  Lorentzian dynamical
effects, for 
in (\ref{eqn:G6})  we can multiply both sides by the rest mass  $m_0$ of the object, and   then
(\ref{eqn:G6}) involves 
\begin{equation}
m({\bf v}_R) =\frac{m_0}{\sqrt{1-\displaystyle{\frac{{\bf v}_R^2}{c^2}}}},
\label{eqn:G8}\end{equation}
the so called `relativistic' mass, and (\ref{eqn:G6}) acquires the form $$\frac{d}{dt}(m({\bf v}_R){\bf
v}_0)={\bf F},$$ where
${\bf F}$ is an effective `force' caused by the inhomogeneities and time-variation of the flow.  This is
essentially Newton's 2nd Law of Motion in the case of gravity only. That $m_0$ cancels is the equivalence principle, 
and which acquires a simple explanation in terms of the flow.  Note that the occurrence of
$1/\sqrt{1-\frac{{\bf v}_R^2}{c^2}}$ will lead to the precession of the perihelion of planetary orbits, and
also to horizon effects wherever  $|{\bf v}| = c$: the region where  $|{\bf v}| < c$ is
inaccessible from the region where $|{\bf v}|>c$. 

Eqn.(\ref{eqn:f4})  involves various absolute quantities such  as the absolute velocity of an object
relative to the  quantum foam and the absolute speed
$c$ also relative to the foam, and of course absolute velocities are excluded from the General Relativity (GR)
formalism.  However (\ref{eqn:f4}) gives (with $t=x_0^0$)
\begin{equation}
d\tau^2=dt^2-\frac{1}{c^2}(d{\bf r}_0(t)-{\bf v}({\bf r}_0(t),t)dt)^2=
g_{\mu\nu}(x_0 )dx^\mu_0 dx^\nu_0,
\label{eqn:24}\end{equation}
which is  the   Panlev\'{e}-Gullstrand form of the metric $g_{\mu\nu}$  \cite{PP, AG} for GR. All of the above
is very suggestive that  useful information for the flow dynamics may be obtained from GR by restricting the
choice of metric to the  Panlev\'{e}-Gullstrand form.  We emphasize that the absolute velocity ${\bf v}_R$ 
has been measured, and so the foundations of GR as usually stated are invalid. As we shall now
see the GR formalism involves two phenomena, namely (i) the use of an unnecessarily  restrictive Einstein
measurement protocol and (ii) the Lorentzian quantum-foam  dynamical effects.  Later we shall remove this
measurement protocol from GR and discover that the GR formalism reduces to explicit fluid flow  equations.   
However to understand  the GR formalism we
need to explicitly introduce the troublesome Einstein  measurement protocol and the peculiar effects that it
induces in the observers historical records.

\vskip12pt
\subsection{  The Einstein Measurement Protocol\label{subsection:theeinsteinmeasurement}}
\vskip6pt

The quantum foam, it is argued, induces actual dynamical time dilations and length contractions in agreement
with the Lorentz interpretation of special relativistic effects.  Then observers in  uniform motion
`through' the foam will on measurement  of the speed of light obtain always the same numerical value  $c$.   To see
this explicitly consider how various observers $P, P^\prime,..$ moving with  different  speeds through the
foam, measure the speed of light.  They  each acquire a standard rod  and an accompanying standardised clock.
That means that these standard  rods  would agree if they were brought together, and at rest with respect to the
quantum foam they would all have length $\Delta l_0$, and similarly for the clocks.    Observer $P$ and
accompanying rod are both moving at  speed $v_R$ relative to the quantum foam, with the rod longitudinal to
that motion. P  then  measures the time
$\Delta t_R$, with the clock at end $A$ of the rod,  for a light pulse to travel from  end $A$ to the other end
$B$  and back again to $A$. The  light  travels at speed $c$ relative to the quantum-foam. Let the time taken for
the light pulse to travel from
$A\rightarrow B$ be $t_{AB}$ and  from $B\rightarrow A$ be $t_{BA}$, as measured by a clock at rest with respect
to the quantum foam\footnote{Not all clocks will behave in this same `ideal' manner.}. The  length of the rod 
moving at speed
$v_R$ is contracted to 
\begin{equation}
\Delta l_R=\Delta l_0\sqrt{1-\frac{v_R^2}{c^2}}.
\label{eqn:c0}\end{equation}
In moving from  $A$ to $B$ the light must travel an extra  distance 
because the  end  $B$ travels a distance $v_Rt_{AB}$ in this time, thus the total distance that must be
traversed  is
\begin{equation}\label{eqn:c1}
ct_{AB}=\Delta l_R+v_Rt_{AB},
\end{equation}
Similarly on returning from $B$ to $A$ the light must travel the distance
\begin{equation}\label{eqn:c2}
ct_{BA}=\Delta l_R-v_Rt_{BA}.
\end{equation}
Hence the total travel time $\Delta t_0$ is
\begin{eqnarray}\label{eqn:c3}
\Delta t_0=t_{AB}+t_{BA}&=&\frac{\Delta l_R}{c-v_R}+\frac{\Delta l_R}{c+v_R}\\
&=&\frac{2\Delta l_0}{c\sqrt{1-\displaystyle\frac{v_R^2}{c^2}}}.
\end{eqnarray}
Because  of  the time dilation effect for the moving clock
\begin{equation}
\Delta t_R=\Delta t_0\sqrt{1-\displaystyle\frac{v_R^2}{c^2}}.
\label{eqn:c4}\end{equation}
Then for the moving observer the speed of light is defined as the distance the observer believes the light
travelled ($2\Delta l_0$) divided by the travel time according to the accompanying clock ($\Delta t_R$), namely 
$2\Delta l_0/\Delta t_R =c$.  So the speed $v_R$ of the observer through the quantum foam  is not revealed by this
procedure, and the observer is erroneously led to the conclusion that the speed of light is always c. 
This follows from two or more observers in manifest relative motion all obtaining the same speed c by this
procedure. Despite this failure  this special effect is actually the basis of the spacetime Einstein measurement
protocol. That this protocol is blind to the absolute motion has led to enormous confusion within physics.

To be explicit the Einstein measurement protocol actually inadvertently uses this special effect by using the radar
method for assigning historical spacetime coordinates to an event: the observer records the time of emission and
reception of radar pulses ($t_r > t_e$) travelling through the space of quantum foam, and then retrospectively
assigns the time and distance of a distant event
$B$ according to (ignoring directional information for simplicity) 
\begin{equation}T_B=\frac{1}{2}(t_r+t_e), \mbox{\ \ \ }
D_B=\frac{c}{2}(t_r-t_e),\label{eqn:25}\end{equation}
  where each observer is now using the same numerical value of $c$.
 The event $B$ is then plotted as a point in 
an individual  geometrical construct by each  observer,  known as a spacetime record, with coordinates $(D_B,T_B)$. This
is no different to an historian recording events according to  some agreed protocol.  Unlike historians, who
don't confuse history books with reality, physicists do so. 
  We now show that because of this
protocol and the quantum foam dynamical effects, observers will discover on comparing their
historical records of the same events that the expression
\begin{equation}
 \tau_{AB}^2 =   T_{AB}^2- \frac{1}{c^2} D_{AB}^2,
\label{eqn:26}\end{equation}
is an invariant, where $T_{AB}=T_A-T_B$ and $D_{AB}=D_A-D_B$ are the differences in times and distances
assigned to events $A$ and
$B$ using the Einstein measurement protocol (\ref{eqn:25}), so long as both are sufficiently small
compared with the scale of inhomogeneities  in the velocity field. 

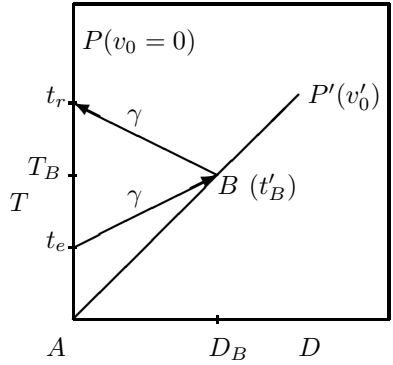
\begin{figure}[ht]
\vspace{-10mm}
\hspace{-7mm}
\setlength{\unitlength}{1.2mm}
\hspace{50mm}\begin{picture}(40,45)
\thicklines
\put(-3,-4){{\bf $A$}}
\put(+1,30){{\bf $P(v_0=0)$}}
\put(16,14){\bf $B$ $(t^\prime_B)$}
\put(25,-4){\bf $D$}
\put(15,-4){\bf $D_B$}
\put(16,-0.5){\line(0,1){0.9}}
\put(-7,12){\bf $T$}
\put(26,24){$ P^\prime(v^\prime_0$)}
\put(0,0){\line(1,0){35}}
\put(0,0){\line(0,1){35}}
\put(0,35){\line(1,0){35}}
\put(35,0){\line(0,1){35}}
\put(0,0){\line(1,1){25}}
\put(0,8){\vector(2,1){16}}
\put(16,16){\vector(-2,1){16}}
\put(-3,8){\bf $t_e$}\put(-0.5,8){\line(1,0){0.9}}
\put(-5,16){\bf $T_B$}\put(-0.5,16){\line(1,0){0.9}}
\put(-3,24){\bf $t_r$}\put(-0.5,24){\line(1,0){0.9}}
\put(6,13){$\gamma$}
\put(6,22){$\gamma$}
\end{picture}
\vspace{5mm}
\caption{\small  Here $T-D$ is the spacetime construct (from  the Einstein measurement protocol) of a special observer
$P$ {\it at rest} wrt the quantum foam, so that $v_0=0$.  Observer $P^\prime$ is moving with speed
$v^\prime_0$ as determined by observer $P$, and therefore with speed $v^\prime_R=v^\prime_0$ wrt the quantum foam. Two light
pulses are shown, each travelling at speed $c$ wrt both $P$ and the quantum foam. As we see later these one-way
speeds for light, relative to the quantum foam, are equal by observation.  Event
$A$ is when the observers pass, and is also used to define zero time  for each for
convenience. }\label{fig:spacetime1}
\end{figure}

To confirm the invariant  nature of the construct in   (\ref{eqn:26}) one must pay careful attention to
observational times as distinct from protocol times and distances, and this must be done separately for each
observer.  This can be tedious.  We now  demonstrate this for the situation illustrated in
Fig.\ref{fig:spacetime1}. 

 By definition  the speed of
$P^\prime$ according to
$P$ is
$v_0^\prime =D_B/T_B$ and so
$v_R^\prime=v^\prime_0$,  where 
$T_B$ and $D_B$ are the protocol time and distance for event $B$ for observer $P$ according to
(\ref{eqn:25}).  Then using (\ref{eqn:26})  $P$ would find that
$(\tau^P_{AB})^2=T_{B}^2-\frac{1}{c^2}D_B^2$ since both
$T_A=0$ and $D_A$=0, and whence $(\tau^{P}_{AB})^2=(1-\frac{v_R^{\prime 2}}{c^2})T_B^2=(t^\prime_B)^2$ where
the last equality follows from the time dilation effect on the $P^\prime$ clock, since $t^\prime_B$ is the time
of event
$B$ according to that clock. Then $T_B$ is also the time that $P^\prime$  would compute for event $B$ when
correcting for the time-dilation effect, as the speed $v^\prime_R$ of $P^\prime$ through the quantum foam is
observable by $P^\prime$.  Then $T_B$ is the `common time' for event $B$ assigned by both
observers\footnote{Because of  gravitational in-flow effects this `common time' is not the same as a
`universal' or `absolute time'; see later. }. For
$P^\prime$ we obtain  directly, also from  (\ref{eqn:25}) and (\ref{eqn:26}), that
$(\tau^{P'}_{AB})^2=(T_B^\prime)^2-\frac{1}{c^2}(D^\prime_B)^2=(t^\prime_B)^2$, as $D^\prime_B=0$  and
$T_B^\prime=t^\prime_B$. Whence for this situation
\begin{equation}
(\tau^{P}_{AB})^2=(\tau^{P'}_{AB})^2,
\label{eqn:invariant1}
\end{equation} and so the
 construction  (\ref{eqn:26})  is an invariant.  

While so far we have only established the invariance of the construct  (\ref{eqn:26}) when one of the
observers is at rest wrt to the quantum foam, it follows that for two observers $P^\prime$ and
$P^{\prime\prime}$ both in motion wrt the quantum foam it follows that they also agree on the invariance
of (\ref{eqn:26}).  This is easily seen by using the intermediate step of  a stationary observer $P$:
\begin{equation}
(\tau^{P'}_{AB})^2=(\tau^{P}_{AB})^2=(\tau^{P''}_{AB})^2.
\label{eqn:invariant2}
\end{equation}
Hence the protocol and Lorentzian effects result in the construction in (\ref{eqn:26})  being indeed an
invariant in general.  This  is  a remarkable and subtle result.  For Einstein this invariance was a
fundamental assumption, but here it is a derived result, but one which is nevertheless deeply misleading.
Explicitly indicating  small quantities  by $\Delta$ prefixes, and on comparing records retrospectively, an
ensemble of nearby observers  agree on the invariant
\begin{equation}
\Delta \tau^2=\Delta T^2-\frac{1}{c^2}\Delta D^2,
\label{eqn:31}\end{equation} 
for any two nearby events.  This implies that their individual patches of spacetime records may be mapped one
into the other merely by a change of coordinates, and that collectively the spacetime patches  of all may
be represented by one pseudo-Riemannian manifold, where the choice of coordinates for this manifold is
arbitrary, and we finally arrive at the invariant 
\begin{equation}
\Delta\tau^2=\eta_{\mu\nu}(x)\Delta x^\mu \Delta x^\nu,
\label{eqn:inv}\end{equation} 
with $x^\mu=\{T,D_1,D_2,D_3\}$ and $\eta_{\mu\nu}$ the usual metric of the spacetime construct.  Eqn.(\ref{eqn:inv})
is of course invariant under the Lorentz transformation.

\vskip12pt
\subsection{ The Origins of General Relativity \label{subsection:theoriginsofgeneral}}
\vskip6pt 

Above it was seen that the Lorentz symmetry of the spacetime construct  would arise if the quantum foam system that
forms space affects the rods and clocks used by observers in the manner indicated.  The effects of absolute motion
with respect to this quantum foam are in fact easily observed, and so the velocity ${\bf v}_R$ of each observer is
measurable.  However if we work only with the spacetime construct then the effects of the absolute motion are
hidden.  Einstein was very much  misled by the reporting of the experiment by Michelson and Morley of 1887, as now
 it is apparent that this experiment, and others since then,  revealed evidence of absolute
motion.  The misunderstanding of the Michelson-Morley experiment had a major
effect on the subsequent development of physics.  One such development was the
work of Hilbert and Einstein  in finding an apparent generalisation of Newtonian
gravity to take into account the apparent absence of absolute motion.  Despite
the deep error in this work the final formulation, known as General Relativity,
has had a number of successes  including the perihelion precession of mercury,
the bending of light and gravitational red shift. Hence despite the incorrect
treatment of absolute motion the formalism of  general relativity  apparently
has some validity.  In the next section we shall {\it deconstruct} this
formalism to discover its underlying physics, but here we first briefly outline
the GR formalism.

 The spacetime construct  is  a static geometrical non-processing historical
record, and is nothing more than a very refined history book, with the shape of the manifold encoded in a metric
tensor $g_{\mu\nu}(x)$.   However in a formal treatment by Einstein the SR formalism and later the GR formalism is
seen to arise from three
 fundamental assumptions:
\begin{eqnarray}
&(1)& \mbox{{\bf The laws of physics have the same form in all }} \nonumber\\
& &    \mbox{{\bf inertial reference frames.}}\nonumber\\
&(2)& \mbox{{\bf Light propagates  through empty space with a  }} \nonumber\\
& & \mbox{{\bf definite  speed c independent of
the speed of the }} \nonumber\\
& & \mbox{{\bf  source or observer.}} \nonumber\\
&(3)& \mbox{{\bf In the limit of low speeds the new formalism }} \nonumber\\
& & \mbox{{\bf should agree with  Newtonian gravity. }}
\label{eqn:EinstPost}\end{eqnarray}

There is strong experimental evidence that all three of these assumptions are in fact wrong (except for the 2nd part
of (2)). Nevertheless  there is something that is partially correct within the formalism, and that part needs to be 
extracted and saved, with the rest discarded. From the above assumptions Hilbert and Einstein
guessed the equation which specifes the  metric tensor $g_{\mu\nu}(x)$, namely the geometry of the spacetime
construct,
\begin{equation}
G_{\mu\nu}\equiv R_{\mu\nu}-\frac{1}{2}Rg_{\mu\nu}=\frac{8\pi G}{c^2} T_{\mu\nu},
\label{eqn:32}\end{equation}
where  $G_{\mu\nu}$ is known as the Einstein tensor, $T_{\mu\nu}$ is the  energy-momentum tensor,
$R_{\mu\nu}=R^\alpha_{\mu\alpha\nu}$ and
$R=g^{\mu\nu}R_{\mu\nu}$ and
$g^{\mu\nu}$ is the matrix inverse of $g_{\mu\nu}$. The curvature tensor is
\begin{equation}
R^\rho_{\mu\sigma\nu}=\Gamma^\rho_{\mu\nu,\sigma}-\Gamma^\rho_{\mu\sigma,\nu}+
\Gamma^\rho_{\alpha\sigma}\Gamma^\alpha_{\mu\nu}-\Gamma^\rho_{\alpha\nu}\Gamma^\alpha_{\mu\sigma},
\label{eqn:curvature}\end{equation}
where $\Gamma^\alpha_{\mu\sigma}$ is the affine connection
\begin{equation}
\Gamma^\alpha_{\mu\sigma}=\frac{1}{2} g^{\alpha\nu}\left(\frac{\partial g_{\nu\mu}}{\partial x^\sigma}+
\frac{\partial g_{\nu\sigma}}{\partial x^\mu}-\frac{\partial g_{\mu\sigma}}{\partial x^\nu} \right).
\label{eqn:affine}\end{equation}
In this formalism the trajectories of test objects are determined by
\begin{equation}
\Gamma^\lambda_{\mu\nu}\frac{dx^\mu}{d\tau}\frac{dx^\nu}{d\tau}+\frac{d^2x^\lambda}{d\tau^2}=0,
\label{eqn:33}\end{equation}
 which is equivalent to minimising the functional
\begin{equation}
\tau[x]=\int dt\sqrt{g^{\mu\nu}\frac{dx^{\mu}}{dt}\frac{dx^{\nu}}{dt}},
\label{eqn:path}\end{equation}
wrt to the path $x[t]$.  

For the case of a spherically symmetric mass a  solution of (\ref{eqn:32}) for $g_{\mu\nu}$ outside of that mass $M$
is the Schwarzschild metric
\begin{equation}
d\tau^2=(1-\frac{2GM}{c^2r})dt^{ 2}-
\frac{1}{c^2}r^{ 2}(d\theta^2+\sin^2(\theta)d\phi^2)-\frac{dr^{ 2}}{c^2(1-\frac{\displaystyle
2GM}{\displaystyle c^2r})}.
\label{eqn:SM}\end{equation}
This solution is the basis of various experimental checks of General Relativity in which the spherically symmetric
mass is either the sun or the earth.  The four tests are: the gravitational redshift, the bending of light, the
precession of the perihelion of mercury, and the time delay of radar signals.

 However the solution (\ref{eqn:SM}) is in fact
completely equivalent to the in-flow interpretation of Newtonian gravity.  Making the change of variables
$t\rightarrow t^\prime$ and
$\bf{r}\rightarrow {\bf r}^\prime= {\bf r}$ with
\begin{equation}
t^\prime=t+
\frac{2}{c}\sqrt{\frac{2GMr}{c^2}}-\frac{4GM}{c^2}\mbox{tanh}^{-1}\sqrt{\frac{2GM}{c^2r}},
\label{eqn:37}\end{equation}
the Schwarzschild solution (\ref{eqn:SM}) takes the form
\begin{equation}
d\tau^2=dt^{\prime 2}-\frac{1}{c^2}(dr^\prime+\sqrt{\frac{2GM}{r^\prime}}dt^\prime)^2-\frac{1}{c^2}r^{\prime
2}(d\theta^{\prime 2}+\sin^2(\theta^\prime)d\phi^{\prime}),
\label{eqn:PG}\end{equation}
which is exactly  the  Panlev\'{e}-Gullstrand form of the metric $g_{\mu\nu}$  \cite{PP, AG} in (\ref{eqn:24})
 with the velocity field given exactly  by the Newtonian form in (\ref{eqn:vfield}).   In which case the trajectory
equation (\ref{eqn:33}) of test objects in the Schwarzschild metric is equivalent to solving (\ref{eqn:G6}). Thus the
minimisation  (\ref{eqn:path}) is equivalent to that of  (\ref{eqn:f4}).  This choice of coordinates corresponds to
a particular frame of reference in which the test object has velocity ${\bf v}_R={\bf v}-{\bf v}_0$ relative to
the in-flow field ${\bf v}$, as seen in (\ref{eqn:f4}).  

 It is conventional wisdom for practitioners in  General
Relativity  to regard the choice of coordinates or frame of reference to be entirely arbitrary and having no physical
significance:  no observations should be possible that can detect and measure ${\bf v}_R$.  This `wisdom' is based
on two  beliefs (i) that all attempts to detect ${\bf v}_R$, namely the detection of absolute motion, have
failed, and that (ii)  the existence of absolute motion is incompatible with the many successes of both the
Special Theory of Relativity and of the General Theory of Relativity.  Both of these beliefs are demonstrably false. 

The results in this section suggest, just as for Newtonian
gravity, that the Einstein General Relativity is nothing more than the dynamical equations for a velocity flow field 
${\bf v}({\bf r },t)$.  Hence  the  spacetime construct appears to be merely an
unnecessary  artifact of the Einstein measurement protocol, which in turn was
motivated by the mis-reporting of the results of the Michelson-Morley
experiment. The successes of General Relativity should thus be considered as an
insight into  the fluid flow dynamics of the quantum foam system, rather than
any confirmation of the validity of the spacetime formalism.  In the next
section we shall deconstruct  General Relativity to extract a possible form for
this dynamics. 

\vskip12pt
\subsection{  Deconstruction of General Relativity
\label{subsection:deconstruction}}
\vskip6pt

Here we deconstruct the formalism of General Relativity by removing the obscurification produced by the
unnecessarily restricted Einstein measurement protocol.  To do this we adopt the  Panlev\'{e}-Gullstrand form  of the
metric
$g_{\mu\nu}$ as that corresponding to the observable quantum foam system, namely to an observationally detected
special frame of reference.  This form for the metric involves a general velocity field ${\bf v}({\bf r},t)$ 
where for precision we consider the coordinates ${\bf r},t$  as that of observers at rest with respect to the CMB
frame.  Note that in this frame  ${\bf v}({\bf r},t)$ is not necessarily zero, for mass acts as a sink for the flow. 
We therefore merely substitute the metric 
\begin{equation}
d\tau^2=g_{\mu\nu}dx^\mu dx^\nu=dt^2-\frac{1}{c^2}(d{\bf r}(t)-{\bf v}({\bf r}(t),t)dt)^2,
\label{eqn:PGmetric}\end{equation}
into (\ref{eqn:32})  using  (\ref{eqn:affine})  and (\ref{eqn:curvature}). This metric involves the arbitrary
time-dependent velocity field  ${\bf v}({\bf r},t)$. This is a very tedious computation and the results below were
obtained  by using the symbolic mathematics capabilities of {\it Mathematica}.  The various components of the
Einstein tensor are then
\begin{eqnarray}\label{eqn:G}
G_{00}&=&\sum_{i,j=1,2,3}v_i\mathcal{G}_{ij}
v_j-c^2\sum_{j=1,2,3}\mathcal{G}_{0j}v_j-c^2\sum_{i=1,2,3}v_i\mathcal{G}_{i0}+c^2\mathcal{G}_{00}, 
\nonumber\\ G_{i0}&=&-\sum_{j=1,2,3}\mathcal{G}_{ij}v_j+c^2\mathcal{G}_{i0},   \mbox{ \ \ \ \ } i=1,2,3.
\nonumber\\ G_{ij}&=&\mathcal{G}_{ij},   \mbox{ \ \ \ \ } i,j=1,2,3.
\end{eqnarray}
where the  $\mathcal{G}_{\mu\nu}$ are  given by
\begin{eqnarray}\label{eqn:GT}
\mathcal{G}_{00}&=&\frac{1}{2}((trD)^2-tr(D^2)), \nonumber\\
\mathcal{G}_{i0}&=&\mathcal{G}_{0i}=-\frac{1}{2}(\nabla\times(\nabla\times{\bf v}))_i,   \mbox{ \ \ \ \ }
i=1,2,3.\nonumber\\ 
\mathcal{G}_{ij}&=&
\frac{d}{dt}(D_{ij}-\delta_{ij}trD)+(D_{ij}-\frac{1}{2}\delta_{ij}trD)trD\nonumber\\ & &
-\frac{1}{2}\delta_{ij}tr(D^2)-(D\Omega-\Omega D)_{ij},  \mbox{ \ \ \ \ } i,j=1,2,3.
\end{eqnarray}
 Here
\begin{equation}
D_{ij}=\frac{1}{2}(\frac{\partial v_i}{\partial x_j}+\frac{\partial v_j}{\partial x_i})
\label{eqn:Dij}\end{equation}
is the symmetric  part of the rate of strain tensor $\frac{\partial v_i}{\partial x_j}$, while the antisymmetric
part is
\begin{equation}
\Omega_{ij}=\frac{1}{2}(\frac{\partial v_i}{\partial x_j}-\frac{\partial v_j}{\partial x_i}).
\label{eqn:Omegaij}\end{equation} 
In vacuum, with $T_{\mu\nu}=0$, we find from (\ref{eqn:32}) and (\ref{eqn:G}) that $G_{\mu\nu}=0$ implies that  
$\mathcal{G}_{\mu\nu}=0$. It is then easy to check that the in-flow velocity field (\ref{eqn:vfield})  satisfies
these equations.  This simply expresses the previous observation that this `Newtonian in-flow' is completely
equivalent to the Schwarzschild metric.  We note that the vacuum equations  $\mathcal{G}_{\mu\nu}=0$ do not involve
the speed of light; it appears only in (\ref{eqn:G}). It is therefore suggested that (\ref{eqn:G}) amounts to the
separation of the Einstein measurement protocol, which involves $c$, from the supposed dynamics of gravity within the  GR
formalism, and which does not involve $c$. However the details of the vacuum dynamics in (\ref{eqn:GT}) have not
actually been tested.  All the key tests of GR are now seen to amount to a test {\it only} of $\delta \tau[x]/\delta
x^\mu = 0$, which is the minimisation of  (\ref{eqn:f4}),  when the in-flow field is given by  (\ref{eqn:G}), and which
is nothing more than Newtonian gravity. Of course Newtonian gravity was itself merely based upon observations within
the solar system, and this may have been too special to have revealed key aspects of gravity. Hence, despite popular
opinion, the GR formalism is apparently based upon  rather poor evidence.

\vskip12pt
\subsection{  The New Theory of Gravity \label{subsection:gravity}}
\vskip6pt
 
Despite the limited insight into gravity which GR is now seen to amount to, here we look for possible
generalisations of Newtonian gravity and its in-flow interpretation by examining some of the mathematical
structures that have arisen in  (\ref{eqn:GT}). For the case of zero vorticity
$\nabla\times{\bf v}=0$ we have
$\Omega_{ij}=0$  and also that we may write 
${\bf v}=\nabla u$ where $u({\bf r},t)$ is a scalar field, and  only one equation is required to determine $u$.
 To that end  we  consider the trace of  
$\mathcal{G}_{ij}$. Note that
$tr(D)=\nabla.{\bf v}$, 
and that
\begin{equation}
\frac{d (\nabla.{\bf v})}{dt}=({\bf v}.\nabla)(\nabla.{\bf v})+\frac{\partial (\nabla.{\bf v})}{\partial t}.
\end{equation}
Then using the identity 
\begin{equation}
({\bf v}.\nabla)(\nabla.{\bf v})=\frac{1}{2}\nabla^2({\bf v}^2)
-tr(D^2)-\frac{1}{2}(\nabla\times{\bf v})^2+ {\bf v}.\nabla\times(\nabla\times{\bf v}),
\label{eqn:identity}\end{equation}
and imposing 
\begin{equation}
\sum_{i=1,2,3}\mathcal{G}_{ii}=-8\pi G\rho,
\label{eqn:new}\end{equation}
 we obtain
\begin{equation}
\frac{\partial}{\partial t}(\nabla.{\bf v})+\frac{1}{2}\nabla^2({\bf v}^2)+\frac{\delta}{4}((tr
D)^2-tr(D^2))=-4\pi G\rho.
\label{eqn:newgravity}\end{equation}
with $\delta=1$. However GR via (\ref{eqn:GT}) also stipulates that $\frac{1}{4}((tr D)^2-tr(D^2))=0$ in vacuum,
implying that  overall $\delta=0$ in GR.  So (\ref{eqn:newgravity}) with $\delta \neq 0$
is not equivalent to GR.  Nevertheless this is seen to be a possible  generalisation of
the Newtonian equation (\ref{eqn:f3}) that includes  the
new term $C({\bf v})=\frac{\delta}{4}((tr D)^2-tr(D^2))$. It appears that the existence
and  significance of this new term has gone unnoticed for some 300 years. 
Its presence explains the many known gravitational anomalies, as we shall see.
Eqn.(\ref{eqn:newgravity}) describes the flow of space and its self-interaction.  The
value of $\delta$   should be determined from both the underlying theory and also by
analysis of experimental data; see Sects.\ref{subsection:measurementsofG} and
\ref{subsection:theborehole}.  We also note that because of the $C({\bf v})$ term  $G$ does not necessarily
have the same value  as the value  $G_N$ determined by say Cavendish type experiments. 

The most significant aspect of (\ref{eqn:newgravity})  is that the new term $C({\bf v})=0$ only for the
in-flow velocity field in (\ref{eqn:vfield}), namely only outside of a spherically symmetric matter
distribution.

Hence (\ref{eqn:newgravity}) in the case of the solar system is indistinguishable from Newtonian
gravity, or the Schwarzschild metric within the General Relativity formalism   so long as we use
(\ref{eqn:f4}), in being able to determine trajectories of test objects.  Hence  (\ref{eqn:newgravity}) is
automatically  in agreement with most of the so-called checks on Newtonian gravity and later General Relativity. 
Note that (\ref{eqn:newgravity}) does not involve the speed of light $c$. Nevertheless  we have not derived 
(\ref{eqn:newgravity})) from the underlying Quantum Homotopic Field Theory, and indeed it is not a consequence of GR, as 
the  $\mathcal{G}_{00}$ equation of  (\ref{eqn:GT}) requires that  $C({\bf v})=0$ in vacuum.
Eqn.(\ref{eqn:newgravity}) at this stage should be regarded as a conjecture  which will  permit the exploration of
possible quantum-flow physics and also allow comparison with experiment. 

However one key aspect of  (\ref{eqn:newgravity}) should be noted
here,  namely that being a non-linear fluid-flow dynamical system we would expect the flow to be
turbulent, particularly when the matter is not spherically symmetric  or inside even a spherically
symmetric distribution of matter, since then the $C({\bf v})$ term is non-zero and it will drive that
turbulence.  We  see that the experiments that reveal absolute motion
also reveal evidence of such turbulence - a new  form of gravitational wave predicted by
the new theory of gravity.

\vskip12pt
\subsection{ The `Dark Matter' Effect \label{subsection:thedark}}
\vskip6pt   

 Because of the  $C({\bf v})$ term  (\ref{eqn:newgravity}) 
would predict that the Newtonian inverse square law would not be applicable  to systems
such as spiral galaxies, because of their highly non-spherical distribution of matter.  Of course
attempts to retain this law, despite its   manifest
 failure,  has   led to the  spurious introduction of  the notion of dark
matter within spiral galaxies, and also at larger scales.  
From 
\begin{equation}\label{eqn:ga}
{\bf g}=\frac{1}{2}\nabla({\bf v}^2)+\frac{\partial {\bf v}}{\partial
t},\end{equation} 
which is (\ref{eqn:f2}) for irrotational flow, we see that (\ref{eqn:newgravity})  gives
\begin{equation}\label{eqn:g2}
\nabla.{\bf g}=-4\pi G\rho-C({\bf v}),
\end{equation}
and taking running time averages to account for turbulence
\begin{equation}\label{eqn:g3}
\nabla.\!\!<\!\!{\bf g}\!\!>=-4\pi G\rho-<\!\!C({\bf v})\!\!>,
\end{equation}
and writing  the extra term as $<\!\!C({\bf v})\!\!>=4\pi G \rho_{DM}$ we see that  $\rho_{DM}$ would act as an
effective matter density, and it is suggested that it is the consequences of this term which have been
misinterpreted as `dark matter'.  Here we see that this effect is actually the consequence of quantum foam
effects within the new proposed dynamics for gravity, and which becomes apparent particularly in spiral
galaxies.  
Because $\nabla\times{\bf v}=0$ we can write (\ref{eqn:newgravity}) in the form

\vspace{4mm}
\noindent${\bf v}({\bf r},t)=$
\begin{equation}
\int^tdt^\prime\int d^3r^\prime ({\bf r}-{\bf r}^\prime)
\frac{
\frac{1}{2}\nabla^2({\bf v}^2({\bf r^\prime},t^\prime))
+4\pi G\rho({\bf r^\prime},t^\prime)
+C({\bf v}({\bf r^\prime},t^\prime))}
{4\pi|{\bf r}-{\bf r}^\prime|^3},
\end{equation}
which allows the  determination of the time evolution of ${\bf v}$.

 In practice it is easier to compute the
 vortex-free velocity field from a velocity potential according to ${\bf v}({\bf r},t)=\nabla u({\bf r},t)$, and we find
the integro-differential equation for $u({\bf r},t)$
\begin{equation}\label{eqn:ueqn}
\frac{\partial u({\bf r},t)}{\partial t}=-\frac{1}{2}(\nabla u({\bf r},t))^2+\frac{1}{4\pi}\int d^3
r^\prime\frac{C(\nabla u({\bf r}^\prime,t))}{|{\bf r}-{\bf r}^\prime|}-\Phi({\bf r},t),
\end{equation}
where $\Phi$ is the Newtonian gravitational potential
\begin{equation}\label{eqn:Phieqn}
\Phi({\bf r},t)=-G\int d^3 r^\prime\frac{\rho({\bf r}^\prime,t)}{|{\bf r}-{\bf r}^\prime|}.
\end{equation}
Hence the  $\Phi$  field acts as the source term for  the velocity potential. Note that in the Newtonian
theory of gravity one has the choice of using either the acceleration field ${\bf g}$ or the velocity field
${\bf v}$. However in the new theory of gravity this choice is no longer available: the fundamental
dynamical degree of freedom is necessarily the ${\bf v}$ field, again because of the presence of the $C({\bf
v})$ term, which obviously cannot be written in terms of ${\bf g}$.

The new flow dynamics encompassed in  (\ref{eqn:newgravity})  thus
accounts for most of the known gravitational phenomena, but will lead to some very clear cut experiments
that will distinguish it from the two previous attempts to model gravitation.  It turns out that these two
attempts were based on some key `accidents' of history. In the case of the Newtonian modelling of gravity
the prime `accident' was of course  the solar system with its high degree of spherical symmetry.  In
each case we had test objects, namely the planets, in orbit about the sun, or we had test object in orbit
about the earth.  In the case of the General Relativity modelling the prime `accident' was the
mis-reporting of the Michelson-Morley experiment, and  the ongoing belief that the so called
`relativistic effects' are incompatible with absolute motion, and of course that GR was constructed to agree with
Newtonian gravity in the `non-relativistic' limit, and so `inherited' the flaws of that theory.  We shall consider in
detail later some further anomalies that might be appropriately explained by this new modelling of gravity.  Of course
that the in-flow has been present in various experimental data is also a significant argument for something like 
(\ref{eqn:newgravity}) to model gravity.

\vskip12pt
\subsection{ In-Flow Superposition Approximation \label{subsection:inflowsuperposition}}
\vskip6pt

We consider here why the existence of absolute motion and as well the consequences and so the presence of the
$C({\bf v})$ term appears to have escaped attention in the case of gravitational experiments and observations
near the earth, despite the fact, in the case of the $C({\bf v})$ term, that the presence of the earth breaks
the spherical symmetry of the matter distribution of   the sun. 

First note that if we have a matter 
distribution $\rho({\bf r})$ at rest in the space of quantum foam, and that (\ref{eqn:newgravity}) has solution
${\bf v}_0({\bf r},t)$,   with ${\bf g}_0({\bf r},t)$ given by (\ref{eqn:ga}), then when the same matter
distribution is uniformly translating at velocity
${\bf V}$, that is 
$\rho({\bf r})\rightarrow  \rho({\bf r}-{\bf V}t)$, then a solution to (\ref{eqn:newgravity}) is 
 \begin{equation}\label{eqn:Vsum}
{\bf v}({\bf r},t)={\bf v}_0({\bf r}-{\bf V}t,t)+{\bf V}.
\end{equation}
Note that this is a manifestly time-dependent process and the time derivative in (\ref{eqn:f2}) or 
(\ref{eqn:G6}) and (\ref{eqn:newgravity}) plays an essential role. As well the result is nontrivial as
(\ref{eqn:newgravity}) is  a non-linear equation. The solution (\ref{eqn:Vsum}) follows because (i) the 
expression for the acceleration${\bf g}({\bf r},t)$ gives, and this expression occurs in (\ref{eqn:newgravity}), 
\begin{eqnarray}\label{eqn:Vsum1}
{\bf g}({\bf r},t) &=&
\frac{\partial {\bf v}_0({\bf r}-{\bf V}t,t)}{\partial t}+(({\bf v}_0({\bf
r}-{\bf V}t,t)+{\bf V}).{\bf \nabla)}({\bf v}_0({\bf r}-{\bf V}t,t)+{\bf
V}),\nonumber\\
&=&
\left.\frac{\partial {\bf v}_0({\bf r}-{\bf V}t^\prime,t)}{\partial t^\prime}\right|_{t^\prime\rightarrow t}+{\bf
g}_0({\bf r}-{\bf V}t,t)+({\bf V}.{\bf \nabla)}{\bf v}_0({\bf r}-{\bf V}t,t),\nonumber\\
&=&
-({\bf
V}.{\bf
\nabla)}{\bf v}_0({\bf r}-{\bf V}t,t)+{\bf g}_0({\bf r}-{\bf V}t,t)+({\bf
V}.{\bf \nabla)}{\bf v}_0({\bf r}-{\bf V}t,t),\nonumber\\
&=&{\bf g}_0({\bf r}-{\bf V}t,t),
\end{eqnarray} 
as there is a key cancellation of two terms in (\ref{eqn:Vsum1}), and (ii) clearly $C({\bf v}_0({\bf
r}-{\bf V}t,t)+{\bf V}) = C({\bf v}_0({\bf r}-{\bf V}t,t))$, and  so this term is  also simply translated.
 Hence apart from the translation effect the acceleration is the same.  Hence the velocity vector
addition rule in (\ref{eqn:Vsum}) is valid for generating the vector flow field for the translating matter
distribution.   This is why the large absolute motion velocity  of some 400 km/s of the solar system
 does not interfere with the usual computation and observation of gravitational forces.

For  earth based  gravitational phenomena the motion of the earth takes place within the velocity in-flow 
towards the sun, and the velocity sum rule (\ref{eqn:Vsum})  is only approximately valid  as now ${\bf V}\rightarrow
{\bf V}({\bf r},t)$ and no longer corresponds to uniform translation, and  manifests turbulence.  To be a valid
approximation the inhomogeneity of ${\bf V}({\bf r},t)$ must be much smaller than that of  ${\bf v}_0({\bf r}-{\bf
V}t,t)$, which it is, as the earth's centripetal acceleration about the sun is approximately 1/1000 that of
the earth's gravitational acceleration at the surface of the earth.  Nevertheless  turbulence   associated  with
the $C({\bf v})$ term is apparent in experimental data.  The validity of this approximation  demonstrates 
that the detection of a cosmic absolute motion and  the in-flow theory of gravity are consistent with the older
methods of computing gravitational forces. This is why both the presence of the $C({\bf v})$ term, the in-flow
and the absolute motion  have gone almost unnoticed in earth based gravitational experiments, except for various
 anomalies; see section \ref{subsection:measurementsofG}.

\vskip12pt
\subsection{ Gravitational In-Flow and the GPS\label{subsection:gps}}
\vskip6pt

It has been  extensively argued that the very successful operation of the Global Positioning System (GPS)
\cite{Ashby} is proof of the validity of the General Relativity formalism for gravity.  However as is well known,
and was most clearly stated by Popper, in science agreement with observation does not amount to the proof of the
theory used to successfully describe the experimental  data; in fact experiment can only strictly be used to
disprove a theory.

We show here that the new in-flow theory of gravity together with the observed absolute velocity of motion of
the solar system through space are together  compatible with the operation of the Global Positioning System
(GPS). Given the developments above  this turns out to be an almost  trivial exercise. 
As usual in this system the effects of the
sun and moon are neglected. Various effects need to be included as the system relies upon extremely accurate atomic
clocks in the satellites forming the GPS constellation.  Within both the new theory and General Relativity
these clocks are affected by both their speed and  the gravitational effects of the earth. As well the orbits
of these satellites and the critical time delays of radio signals from the satellites need to be computed.  For the
moment we assume spherical symmetry for the earth. The effects of non-sphericity will be discussed below. 
In General Relativity the orbits and signalling time delays are determined by the use of the geodesic equation
(\ref{eqn:33}) and the Schwarzschild metric (\ref{eqn:SM}).  However these two equations are equivalent to
the  orbital equation (\ref{eqn:G8}) and the velocity field  (\ref{eqn:Vsum}), with a velocity ${\bf V}$ of
absolute motion, and  with the in-flow given by (\ref{eqn:vfield}), noting the result in
section \ref{subsection:inflowsuperposition}. For EM signalling the elapsed time in (\ref{eqn:f4})
requires careful treatment.  Hence the two systems are completely mathematically equivalent:  the
computations within the new system may most easily be considered by relating them to the mathematically
equivalent General Relativity formalism.  
We can also see this by explicitly changing from the CMB frame to a non-rotating frame co-moving with the earth
by means of  the change of variables
\begin{eqnarray}\label{eqn:frame1}
{\bf r}&=&{\bf r}^\prime+{\bf V}t^\prime,  \\ \nonumber
t&=&t^\prime, \\  \nonumber
{\bf v}&=&{\bf v}^\prime+{\bf V},
\end{eqnarray}  
which lead to  the relationships of differentials
\begin{eqnarray}\label{eqn:frame2}
\nabla^\prime &=& \nabla, \\  \nonumber
\frac{\partial}{\partial t ^\prime}&=&\frac{\partial}{\partial t}+{\bf V}.\nabla
\end{eqnarray}
These expressions then lead to the demonstration of the invariance of (\ref{eqn:newgravity}). Then in the earth
co-moving frame the absolute velocity ${\bf V}$ does not appear in (\ref{eqn:newgravity}).  Then another change of
variables, as in  (\ref{eqn:37}), permits (\ref{eqn:newgravity}) to be written in the form of General Relativity
with a Schwarzschild metric.

The consistency between the absolute motion velocity ${\bf V}$ and General Relativity may also be directly checked by showing
explicitly, using say {\it Mathematica}, that the metric
\begin{equation}
d\tau^2=g_{\mu\nu}dx^\mu dx^\nu=dt^2-\frac{1}{c^2}(d{\bf r}(t)-(({\bf v}({\bf r}-{\bf V}t)+{\bf V})dt)^2,
\label{eqn:PGmetric2}\end{equation}
is a solution to (\ref{eqn:32}) for $T_{\mu\nu}=0$, ie outside matter, where ${\bf v}({\bf r})$ is the in-flow velocity field in 
(\ref{eqn:vfield}). This metric is a generalisation of the Panlev\'{e}-Gullstrand metric to include the absolute motion effect. This
emphasises yet again that for a spherically symmetric matter distribution the Schwarzschild metric, which is equivalent to the
Panlev\'{e}-Gullstrand metric, is physically identical to  Newtonian gravity.

   There are nevertheless two  differences between the two theories. One is their different
treatment of the non-sphericity of the earth  via the $C({\bf v})$ term, and the second difference is the effects
of the in-flow turbulence.  In the operation of the GPS the density $\rho({\bf r})$ of the earth is not used. 
Rather the gravitational potential $\Phi({\bf r})$ is determined observationally.  In the new gravity theory the
determination of such a gravitational potential via (\ref{eqn:newgravity})  and 
$\Phi({\bf r})=-\frac{1}{2}{\bf v}^2({\bf r})$ would involve the extra  $C({\bf v})$ term.  Hence because of
this phenomenological treatment the effects of the  $C({\bf v})$ term are not checkable.   However the
gravitational wave effect is expected to affect the operation of the GPS, and the GPS constellation
would offer a worldwide network  which would enable the investigation of the spatial and temporal correlations
of these gravitational waves. 

There is also a significant interpretational difference between the two theories. For example in  General
Relativity the relativistic effects involve both the `special relativity' orbital speed  effect via time
dilations of the satellite clocks together with the General Relativity `gravitational potential energy' effect on
the satellite clocks. In the new theory there is only one effect, namely the time dilation effect produced by the
motion of the clocks through the quantum foam, and the speeds of these clocks involve the vector sum of the
orbital velocity and the velocity caused by the in-flow of the quantum foam into the earth. 

The relations in (\ref{eqn:frame2}) are those of Galilean Relativity.  However together with these go the
real absolute motion effects of time dilations and length contractions for moving material systems. Then the data
from  observers in absolute motion may be related by the Lorentz transformation, so long as their data is not
corrected for the effects of absolute motion.  So the new Process Physics brings together transformations that
were, in the past, regarded as mutually exclusive.

\vspace{3mm}

\vskip12pt
\subsection{  Measurements of $G$ \label{subsection:measurementsofG}}

\begin{figure}[t]
\hspace{20mm}\includegraphics[scale=1.0]{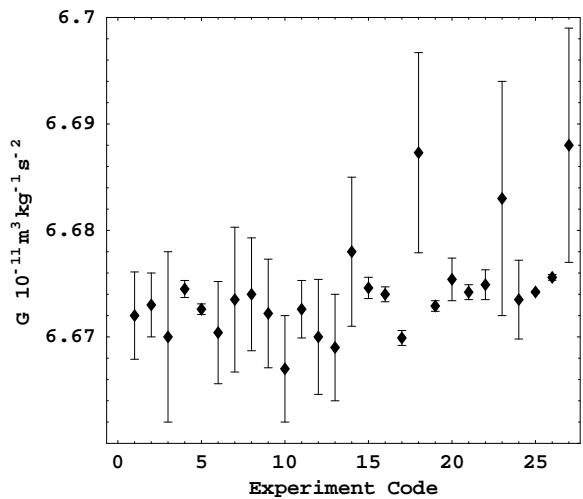}
\caption{\small{Results of precision measurements of $G$ published in the 
last sixty years in which the Newtonian theory was used to analyse the data.  These results show  the presence of
a  systematic effect not in the Newtonian theory. {\bf 1:}  Gaithersburg 1942 \cite{Gaithersburg42},
{\bf 2:}  Magny-les-Hameaux 1971 \cite{Magny-les-Hameaux}, 
{\bf 3:}  Budapest 1974   \cite{Budapest}, 
{\bf 4:}  Moscow 1979  \cite{Moscow79},
{\bf 5:}  Gaithersburg 1982 \cite{Gaithersburg82}, 
{\bf 6-9:}  Fribourg   Oct 84, Nov 84, Dec 84, Feb 85 \cite{Fribourg},
{\bf 10:}    Braunschweig 1987   \cite{Braunschweig87},
{\bf 11:}    Dye 3 Greenland   1995       \cite{Dye3},
{\bf 12:}    Gigerwald Lake 1994   \cite{Gigerwaldlake94}, 
{\bf 13-14:} Gigerwald lake19 95  112m, 88m     \cite{Gigerwaldlake95},
{\bf 15:}     Lower Hutt 1995    MSL  \cite{LowerHutt95}, 
{\bf 16:}  Los Alamos 1997 \cite{Los Alamos}, 
{\bf 17:}    Wuhan 1998  \cite{Wuhan},
{\bf 18:}    Boulder JILA 1998  \cite{Boulder}, 
{\bf 19:}    Moscow 1998  \cite{Moscow98}, 
{\bf 20:}    Zurich 1998  \cite{Zurich98}, 
{\bf 21:}     Lower Hutt MSL 1999   \cite{LowerHutt99}, 
{\bf 22:}    Zurich 1999   \cite{Zurich99}, 
{\bf 23:}    Sevres 1999  \cite{Sevres99}, 
{\bf 24:}   Wuppertal 1999  \cite{Wuppertal},
{\bf 25:}     Seattle 2000 \cite{Seattle},
{\bf 26:}     Sevres 2001  \cite{Sevres01}, 
{\bf 27:}     Lake Brasimone 2001 \cite{lake Brasimone}. 
  Data compilation  adapted from \cite{MMGdata}. }   
\label{fig:GData}}\end{figure}
\vskip6pt

As noted in Sect.\ref{subsection:classicalquantum} Newton's Inverse Square Law of Gravitation
is strictly valid only in cases of spherical symmetry, and then only outside of such a matter distribution.  The 
theory that gravitational effects arise from inhomogeneities in the quantum foam flow  implies that there is no
`universal law of gravitation' because the inhomogeneities are determined by non-linear `fluid equations' and the
solutions  have no form which could be described by a `universal law'.  Fundamentally there is no
generic fluid flow behaviour. The Inverse Square Law is then only an approximation, with  large
deviations expected in the case of spiral galaxies. Nevertheless Newton's gravitational constant $G$
will have a definite value as it quantifies the effective rate at which matter dissipates the
information content of space.  

From these considerations it
follows that the measurement of the value of $G$ will be difficult as the measurement of the  forces
between two of more objects, which is the usual method of measuring $G$, will depend on the geometry
of the spatial positioning of these objects  in a way not previously accounted for because the
Newtonian Inverse Square Law has always been assumed, or in some cases  a specified change in the
form of the law has been used.  But in all cases a `law' has been assumed, and this may have been
the flaw in  the analysis of data from such experiments.  This implies that the value of
$G$ from such experiments will show some variability as a systematic effect has  been neglected in
analysing the experimental data, for in none of these experiments is spherical symmetry present.  So
experimental measurements of $G$  should show an unexpected contextuality.  As well the influence of
surrounding matter has also not been properly accounted for. Of course  any effects of turbulence in
the inhomogeneities of the flow has presumably never even been contemplated.  

The first measurement of $G$ was in 1798 by Cavendish using a torsional balance.  As the precision of
experiments increased over the years and a variety of techniques used  the disparity between the
values of $G$ has actually increased, as shown in Fig.\ref{fig:GData}, and as reviewed in \cite{Gillies}.  In 1998
CODATA increased the uncertainty in
$G$ from 0.013\% to 0.15\%.  One indication of the contextuality is that measurements of $G$  produce values that
differ by nearly 40 times their individual error estimates .  It is predicted that these $G$ anomalies will only be
resolved when  the new theory of gravity is used in analysing the data from these experiments.

\vskip12pt
\subsection{  Gravitational Anomalies\label{subsection:gravitationalanomalies}}
\vskip6pt

In Sect.\ref{subsection:measurementsofG} anomalies associated with the measurement of $G$ were briefly discussed and
it was pointed out that these were probably explainable within the new in-flow theory of gravity.  There are in-fact
additional gravitational anomalies that are not well-known in physics, presumably because their existence is
incompatible with the Newtonian or the  Hilbert-Einstein gravity theories.

The most significant of these anomalies is the Allais effect \cite{Allais}.   In June 1954 Allais\footnote{Maurice
Allais won the Noble Prize for Economics in 1988.} reported that a Foucault
pendulum exhibited peculiar movements at the time of a solar eclipse.  Allais was recording the precession of a
Foucault pendulum in Paris. Coincidently during the 30 day observation period a partial solar eclipse occurred at
Paris on June 30.  During the eclipse the precession of the pendulum was seen to be disturbed.  Similar results were
obtained during another solar eclipse on October 29 1959.  There have been other repeats of the Allais experiment
with varying results.  

Another anomaly was reported by Saxl and Allen \cite{Saxl} during the solar eclipse of March 7 1970.  Significant
variations in the period of a torsional pendulum were observed  both during the eclipse and as well in the hours just
preceding and just following the eclipse.  The effects seem  to suggest that an ``apparent wavelike structure has been observed over the
course of many years at our Harvard laboratory'', where the wavelike structure is present and reproducible even in the
absence of an eclipse. 

Again Zhou and Huang \cite{Zhou1,Zhou2,Zhou3}  report various time anomalies occurring during the solar eclipses of
September 23 1987,  March 18 1988 and  July 22 1990 observed using atomic clocks.

Another anomaly is associated with the rotational velocities of objects in spiral galaxies, which are larger than
could be maintained by the apparent amount of matter in such galaxies. This anomaly led to the introduction of
the `dark matter' concept - but with no such matter ever having been detected, despite extensive searches. This
anomaly was compounded when recently observations of the rotational velocities of objects within elliptical
galaxies was seen to require very little `dark matter' \cite{Elliptical}.  Of course this is a simple consequence of
the new theory of gravity.  The `dark matter' effect is nothing more than an aspect of the self-interaction of space
that is absent in both the Newtonian and General Relativity theories.  As a system becomes closer to being spherically
symmetric, such as in the transition from spiral to elliptical galaxies, the new $C({\bf v})$ term  becomes less
effective.

All these anomalies, including the $g$ anomaly in sect.\ref{subsection:theborehole}, and others such as the
Pioneer 10/11 de-acceleration anomaly \cite{Anderson} and the solar neutrino flux deficiency problem, not discussed
here, would suggest that  gravity has aspects to it that are not within the prevailing theories, but that the in-flow
theory discussed above might well provide an explanation, and indeed these anomalies may well provide further phenomena
that could be used to test the new theory.  The effects  associated with the solar eclipses could presumably
follow from the alignment of the sun, moon and the earth causing enhanced turbulence.  The Saxl and Allen
experiment of course suggests, like the other experiments, that the turbulence is always present. To explore
these anomalies detailed numerical studies of (\ref{eqn:newgravity}) are required with particular emphasis on the
effect on the position of the moon.

\vskip12pt
\subsection{  The Borehole g Anomaly\label{subsection:theborehole}}
\vskip6pt

\begin{figure}[ht]
\hspace{35mm}\includegraphics[scale=0.7]{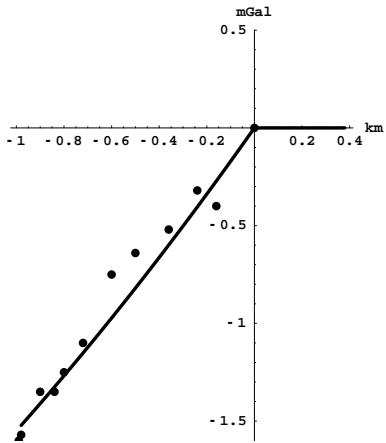}
\caption{\small{ The data shows the gravity residuals for the Hilton mine profile, from Ref.\cite{Stacey2},  defined as
$\Delta g(r) = g_{Newton}-g_{observed}$, and measured in mGal (1mGal $ =10^{-3}$ cm/s$^2$) plotted against depth in km. The theory curve shows
$\Delta g(r) = g_{Newton}-g_{InFlow}$ from solving (\ref{eqn:NewtRadial}) and (\ref{eqn:InFlowRadial})  for a 
density
$\rho = 2760$ kg/m$^3$ appropriate to the Hilton mine,  a coefficient
$\delta = 1$ and $G=0.99925 G_N$. } 
\label{fig:gAnomaly}}\end{figure}

Stacey and others \cite{Stacey1,HST86,Stacey2} have found evidence for non-Newtonian gravitation from
gravimetric measurements (Airy experiments) in mines and boreholes.  The discovery was that the measured
value of $g$ down mines and boreholes became greater than that predicted by the Newtonian theory, given the
density profile
$\rho(r)$ implied by sampling, and so implying a defect in Newtonian gravity, as shown in Fig.\ref{fig:gAnomaly} for
the Hilton mine.  The results were interpreted and analysed using either a value of $G$  different to but larger
than that found in laboratory experiments or by assuming a short range Yukawa type force in addition to the Newtonian
`inverse-square law'.  Numerous experiments were carried out in which
$g$ was measured as a function of depth, and also as a function of height above ground level using towers. The
tower experiments \cite{Thomas89,Jekeli} did not indicate any non-Newtonian effect, and so implied that the
extra Yukawa force explanation was not viable.   The combined results appeared to have resulted in
confusion and eventually the experimental effect was dismissed as being caused by erroneous
density sampling \cite{Forces}.  However the new theory of gravity predicts such an effect, and in particular that the
effect should manifest within the earth but not above it, as was in fact observed.    Essentially this effect is
caused by the new $C({\bf v})$ term in the in-flow theory of gravity which, as we have noted earlier, is
active whenever there is a lack of complete spherical symmetry, or even within matter when there is
spherical symmetry - this being the case here.

The Newtonian in-flow equation (\ref{eqn:f3}) for a time-independent velocity field becomes  for systems with spherical
symmetry
\begin{equation} 
2\frac{vv^\prime}{r} +(v^\prime)^2 + vv^{\prime\prime}=-4\pi\rho(r)G_N, 
\label{eqn:NewtRadial}
\end{equation}
where $ v=v(r)$ and $v^\prime= \frac{d v(r)}{dr}$. The value of $v$ at the earth's surface is approximately
$11$ km/s. This formulation is completely equivalent to
the conventional formulation of Newtonian gravity,

In the new  gravity theory the in-flow equation (\ref{eqn:newgravity}) has the additional $C({\bf v})$  term
which, in the case of time-independent  flows and spherical symmetry, becomes the term in the brackets in
(\ref{eqn:InFlowRadial}) with coefficient $\delta$, 

\begin{equation}
2\frac{vv^\prime}{r} +(v^\prime)^2 + vv^{\prime\prime} + \delta(\frac{v^2}{2r^2}+
\frac{vv^\prime}{r})=-4\pi\rho(r)G. 
\label{eqn:InFlowRadial}
\end{equation}
It is important to note that the value of $G$ is not necessarily the same as the conventional
value denoted as $G_N$.
Both of these equations may be integrated in from the surface, assuming that the in-flow velocity field at or above 
the surface is given by
\begin{equation}
 v(r)=\sqrt{\frac{2G_N M}{r}},
\label{eqn:vin}\end{equation}
so that it corresponds to the observed surface value of $g$.  In (\ref{eqn:NewtRadial})  $M$ is the total
matter content of the earth, but in (\ref{eqn:InFlowRadial}) $M$ is the sum of the matter content and the
effective total  `dark matter' content of the earth. Then above the surface, where
$\rho=0$, both flow equations have (\ref{eqn:vin}) as identical solutions, since for this velocity field the
additional bracketed term in (\ref{eqn:InFlowRadial}) is identically zero. This explains why the tower
experiments found no non-Newtonian effects.  The in-flow equations may be numerically integrated inward
from the surface using as boundary conditions the continuity of $v(r)$ and $v^\prime(r)$ at the surface. 
For each the
$g(r)$ is determined.  Fig.\ref{fig:gAnomaly}  shows the resulting difference  $\Delta g(r) =
g_{Newton}-g_{InFlow}$ compared with the measured anomaly  $\Delta g(r) = g_{Newton}-g_{observed}$.  Assuming
$\delta=1$ the value of $G$ was adjusted to agree with the data, giving $G=0.99925 G_N$, as shown in
Fig.\ref{fig:gAnomaly}. However this fit does not uniquely determine the values of $\delta$ and $G$.  It should be
noted that the data in Fig.\ref{fig:gAnomaly}  was adjusted for density irregularities using Newtonian gravity, and this is now
seen to be an invalid procedure.  Nevertheless  the results imply that a repeat of the borehole measurements would be very
useful in contributing to the testing of the new theory of gravity, or perhaps even a re-analysis of existing data could be
possible.  The key signature of the effect, as shown in  Fig.\ref{fig:gAnomaly}, is the discontinuity in $d\Delta g(r)/dr$ at
the surface, and which is a consequence of  having  the $C({\bf v})$ term. Of course using a Yukawa force added to Newtonian
gravity cannot produce this key signature, as such a force results in $d\Delta g(r)/dr$ being continuous at the surface.

\vskip12pt
\section{ Detection of Absolute Motion  and Gravitational Waves\label{section:detectionofabsolute}}
\vskip6pt

\vskip12pt
\subsection{Space and Absolute Motion\label{subsection:spaceandabsolute}}
\vskip6pt

Absolute motion is motion relative to space itself.  It turns out that Michelson and Morley in their historic
experiment of 1887 did detect absolute motion, but rejected their own findings because using their method of
analysis of the observed fringe shifts the determined speed of some 8 km/s was less than the 30 km/s orbital speed of
the earth.  The data was clearly indicating that the theory for the operation of the Michelson interferometer was not
adequate.  Rather than reaching this conclusion Michelson and Morley came to the incorrect conclusion that their
results amounted to the failure to detect absolute motion.   This had an enormous impact on the development of
physics, for as is well known Einstein adopted the absence  of absolute motion effects as one of his fundamental
assumptions.  By the time Miller had finally figured out how to work around the lack of a viable theory for the
operation of the Michelson interferometer, and properly analyse data from his own Michelson interferometer,
absolute motion had become a forbidden concept within physics, as it still is at present.  The  experimental
observations  by Miller and others of absolute motion has continued to be scorned and rejected by the physics
community.  Fortunately as well as revealing absolute motion the experimental data also reveals evidence in
support of a new theory of gravity.   

\vskip12pt
\subsection{Theory of the  Michelson Interferometer\label{subsection:theoryof}}
\vskip6pt

\begin{figure}[h]

\vspace{3mm}
\setlength{\unitlength}{0.8mm}
\hspace{28mm}\begin{picture}(0,30)
\thicklines
\put(-10,0){\line(1,0){50}}
\put(-5,0){\vector(1,0){5}}
\put(40,-1){\line(-1,0){29.2}}
\put(15,0){\vector(1,0){5}}
\put(30,-1){\vector(-1,0){5}}
\put(10,0){\line(0,1){30}}
\put(10,5){\vector(0,1){5}}
\put(11,25){\vector(0,-1){5}}
\put(11,30){\line(0,-1){38}}
\put(11,-2){\vector(0,-1){5}}
\put(8.0,-2){\line(1,1){5}}
\put(9.0,-2.9){\line(1,1){5}}
\put(6.5,30){\line(1,0){8}}
\put(40,-4.5){\line(0,1){8}}
\put(5,12){ $L$}
\put(4,-5){ $A$}
\put(35,-5){ $B$}
\put(25,-5){ $L$}
\put(12,26){ $C$}
\put(9,-8){\line(1,0){5}}
\put(9,-9){\line(1,0){5}}
\put(14,-9){\line(0,1){1}}
\put(9,-9){\line(0,1){1}}
\put(15,-9){ $D$}
\put(50,0){\line(1,0){50}}
\put(55,0){\vector(1,0){5}}
\put(73,0){\vector(1,0){5}}
\put(85,0){\vector(1,0){5}}
\put(90,15){\vector(1,0){5}}
\put(100,-1){\vector(-1,0){5}}
\put(100,-4.5){\line(0,1){8}}
\put(68.5,-1.5){\line(1,1){4}}
\put(69.3,-2.0){\line(1,1){4}}
\put(70,0){\line(1,4){7.5}}
\put(70,0){\vector(1,4){3.5}}
\put(77.5,30){\line(1,-4){7.7}}
\put(77.5,30){\vector(1,-4){5}}
\put(73.5,30){\line(1,0){8}}
\put(83.3,-1.5){\line(1,1){4}}
\put(84.0,-2.0){\line(1,1){4}}
\put(100,-1){\line(-1,0){14.9}}
\put(67,-5){ $A_1$}
\put(82,-5){ $A_2$}
\put(95,-5){ $B$}
\put(79,26){ $C$}
\put(90,16){ $v$}
\put(-8,8){(a)}
\put(55,8){(b)}

\end{picture}

\vspace{10mm}
\caption{\small{Schematic diagrams of the
Michelson Interferometer, with
beamsplitter/mirror at $A$ and mirrors at $B$ and
$C$, on equal length arms when parallel, from $A$. $D$ is a quantum detector (not drawn in (b)) that causes
localisation  of the photon state by a collapse process. In (a) the interferometer is at
rest in space. In (b) the interferometer is moving with speed $v$ relative to space in
the direction indicated. Interference fringes are observed at the quantum detector $D$. 
If the interferometer is rotated in the plane  through $90^o$, the roles of arms
$AC$ and $AB$ are interchanged, and during the rotation shifts of the fringes are seen
in the case of absolute motion, but only if the apparatus operates in a gas.  By counting
fringe changes the speed $v$ may be determined.}\label{fig:Minterferometer}}
\end{figure}
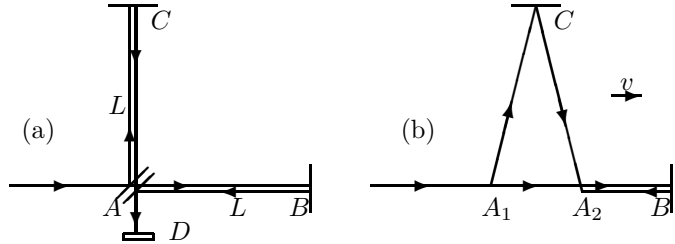

We now show for the first time in over 100 years how   three key effects together permit the 
Michelson interferometer \cite{Mich} to  reveal the phenomenon of absolute
motion when operating in the presence of a gas, with the third effect only discovered in 2002
\cite{CK}.       The main outcome is the  derivation of the origin of the Miller
$k^2$  factor in the expression for the time difference for light travelling via the orthogonal arms,
\begin{equation}\label{eqn:QG0}
\Delta t=k^2\frac{L|{\bf v}_P|^2}{c^3}\cos(2(\theta-\psi)).
\end{equation}
 Here ${\bf v}_P$ is the projection of the absolute velocity ${\bf v}$ of the
interferometer  through the quantum-foam  onto the plane of the interferometer, where the projected
velocity vector ${\bf v}_P$ has azimuth angle $\psi$ relative to the local meridian, and
$\theta$ is the angle of one arm from that meridian.  The  $k^2$ factor is   $k^2=n(n^2-1)$  where $n$
is the refractive index of the gas through which the light  passes,  $L$ is the
length of each arm and
$c$ is the speed of light relative to the quantum foam. This expression follows from three key effects: (i)
the difference in geometrical length of the two paths when the interferometer is in absolute motion, as
first realised by Michelson,  (ii) the Fitzgerald-Lorentz contraction of the arms along the direction of
motion, and (iii) that these two effects precisely cancel in vacuum, but leave a residual effect if
operated in a gas,  because the speed of light through the gas is reduced to
$V=c/n$, ignoring here for simplicity any  Fresnel-drag effects. 
This is one of the  aspects of the
quantum foam physics that distinguishes it from the Einstein formalism.  The time difference
$\Delta t$ is revealed by the fringe shifts on rotating the interferometer. In Newtonian physics,
that is with no Fitzgerald-Lorentz contraction, $k^2=n^3$, while in Einsteinian physics $k=0$
reflecting the fundamental assumption that absolute motion is not measurable and indeed has no
meaning.   So the experimentally determined value of
$k$ is a key test of fundamental physics.    For  air $n=1.00029$, and so for
process physics $k=0.0241$ and $k^2=0.00058$, which is close to the Einsteinian value of $k=0$,
particularly in comparison to the Newtonian value of $k=1.0$.  This small but non-zero $k$ value explains
why the Michelson interferometer experiments gave such small fringe shifts.  Fortunately it is possible to
check  the $n$ dependence of $k$ as two experiments \cite{Illingworth, Joos} were done in helium gas, and this has
an $n^2-1$ value significantly different from that of air.

In deriving  (\ref{eqn:QG1}) in the new physics it is essential to note that space is a quantum-foam system
  which exhibits various subtle features. In particular it exhibits real dynamical
effects on clocks and rods. In this physics the speed of light is only $c$  relative to the
quantum-foam, but to observers moving with respect to this quantum-foam the speed appears to be still $c$, but
only because their clocks and rods are affected by the quantum-foam. As shown in 
above such observers will find that records of observations of
distant events will be described by the Einstein spacetime formalism, but only if they restrict
measurements to those achieved by using clocks, rods and light pulses, that is using the Einstein
measurement protocol.   However if they use an absolute motion detector then such observers can correct
for these effects. 

It is simplest
in the new physics to work in the quantum-foam frame of reference.  If there is a gas present at rest in
this frame, such as air, then the speed of light in this frame is
$V=c/n$. If the interferometer and gas  are moving with respect to the quantum foam, as in the case of an
interferometer attached to the earth, then the speed of light relative to the quantum-foam is still
$V=c/n$ up to corrections due to  drag effects.    Hence this new physics requires a different method of analysis
from that of the Einstein physics. With these cautions we now describe the operation of a Michelson
interferometer in this new physics, and show that it makes predictions different to that of the Einstein
physics.    Of course experimental evidence is the final arbiter in this conflict of theories.

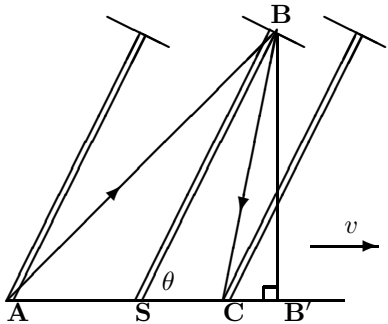
\begin{figure}[ht]
\vspace{15mm}\setlength{\unitlength}{0.9mm}
\hspace{35mm}\begin{picture}(20,30)
\thicklines
\put(0,0){\line(1,0){50}}
\put(0,-3){{\bf A}}
\put(32,0){\line(1,5){8}}

\put(32,-3){{\bf C}}
\put(19,0){\line(1,2){20.0}}
\put(20,0){\line(1,2){20}}
\put(19,-3){{\bf S}}
\put(23,+1.5){{$\theta$}}

\put(35,42){\line(2,-1){9}}
\put(39,41){{\bf B}}
\put(0,0){\line(1,1){39.6}}
\put(50,10){{$ v$}}
\put(45,8){\vector(1,0){10}}

\put(0,0){\line(1,2){19.8}}
\put(1,0){\line(1,2){19.6}}
\put(15,42){\line(2,-1){9}}

\put(32,0){\line(1,2){19.8}}
\put(33,0){\line(1,2){19.6}}
\put(47,42){\line(2,-1){9}}

\put(40,0){\line(0,1){39.5}}
\put(41,-3){{\bf B}$'$}

\put(38,0){\line(0,1){2}}
\put(38,2){\line(1,0){2}}

\put(15,15){\vector(1,1){2}}
\put(35.02,15){\vector(-1,-4){0.5}}

\end{picture}

\vspace{3mm}
\caption{\small{One arm  of a Michelson Interferometer travelling  at  angle $\theta$ and   velocity 
${\bf v}$, and shown at three successive times: (i) when photon leaves beamsplitter at $A$, (ii) when photon
is reflected at mirror $B$, and (iii) when photon returns to beamsplitter at $C$. The line $BB'$ defines
right angle triangles $ABB'$ and $SBB'$.  The second arm is not shown but has angle $\theta+90^o$ to 
${\bf v}$. Here ${\bf v}$ is in the plane of the interferometer for simplicity, and
the azimuth angle $\psi=0$. }\label{fig:MMangled}}
\end{figure}

As shown in Fig.\ref{fig:MMangled}  the  beamsplitter/mirror when  at $A$ sends a photon $\psi(t)$ into a
superposition
$\psi(t)=\psi_1(t)+\psi_2(t)$, with each component travelling in different arms of the interferometer, until
they are recombined in the quantum detector which results in a localisation process, and one spot in the
detector is produced.  Repeating with many photons reveals that the
interference between
$\psi_1$ and
$\psi_2$ at the detector results in fringes.  These fringes actually only appear if the mirrors are not quite
orthogonal, otherwise the screen has  a uniform intensity and this intensity changes as the interferometer
is rotated, as shown in the analysis by  Hicks \cite{Hicks}.   To simplify the analysis here assume that
the two arms are constructed to have the same lengths
$L$  when they are physically parallel to each other and perpendicular to $ v$, so that the distance 
$BB'$ is
$L\sin(\theta)$. The Fitzgerald-Lorentz effect in the new physics  is that the distance  $SB'$  is 
$\gamma^{-1} L\cos(\theta)$ where
$\gamma=1/\sqrt{1-v^2/c^2}$.  The various other distances  are $AB=Vt_{AB}$, $BC=Vt_{BC}$, $AS=vt_{AB}$ 
and $SC=vt_{BC}$, where $t_{AB}$ and $t_{BC}$ are the travel times.  Applying the Pythagoras theorem to
triangle $ABB'$ we obtain
\begin{eqnarray}\label{eqn:QG1}
t_{AB}&=&\frac{2v\gamma^{-1}
L\cos(\theta)}{2(V^2-v^2)}+\nonumber \\
&
&\frac{\sqrt{4v^2\gamma^{-2}L^2\cos^2(\theta)+4L^2(1-\frac{v^2}{c^2}\cos^2
(\theta))(V^2-v^2)}}{2(V^2-v^2)}.\nonumber\\
\end{eqnarray}

\noindent The expression for $t_{BC}$ is the same except for a change of sign of the  $2v\gamma^{-1}
L\cos(\theta)$ term, then  
\begin{eqnarray}\label{eqn:QG2}  
t_{ABC}&=&t_{AB}+t_{BC}\nonumber \\
&=&\frac{\sqrt{4v^2\gamma^{-2}L^2\cos^2(\theta)+4L^2(1-\frac{v^2}{c^2}\cos^2(\theta))
(V^2-v^2)}}{(V^2-v^2)}.\nonumber \\
\end{eqnarray}
The  corresponding travel time $t'_{ABC}$ for the orthogonal arm  is obtained from  (\ref{eqn:QG2}) by the
substitution $\cos(\theta) \rightarrow \cos(\theta+90^0)=-\sin(\theta)$. The difference in travel times
between the two arms is then $\Delta t= t_{ABC}-t'_{ABC}$. Now trivially $\Delta t =0$  if $v=0$, but 
also $\Delta t =0$ when
$v\neq 0$ but only if $V=c$.  This then would  result in a null result on rotating the apparatus.  Hence the null
result of  Michelson interferometer  experiments in the new physics is only for the special case of
photons travelling in vacuum for which $V=c$.    However if the interferometer is immersed
 in a gas then $V<c$ and a non-null effect is expected on rotating the apparatus, since now 
$\Delta t \neq 0$.  It is essential then in analysing data to correct for this refractive index effect.  The above
$\Delta t$ is the change in travel time when one arm is moved through angle $\theta$.  The interferometer
operates by comparing the change in the difference of the travel times between the arms.  Then for
$V=c/n$ we  find for $v << V$  that 
\begin{equation}\label{eqn:QG4}
\Delta t= Ln(n^2-1)\frac{v^2}{c^3}\cos(2\theta)+\mbox{O}(v^4),
\end{equation}
that is $k^2=n(n^2-1)$,  which gives $k=0$ for vacuum experiments ($n=1$).  So the Miller
phenomenological parameter $k$  is seen to accommodate  both the Fitzgerald-Lorentz
contraction effect and the dielectric effect, at least for gases.  This is very fortunate since being
a multiplicative parameter a re-scaling of old analyses is all that is required. $\Delta t$ is
non-zero  when
$n \neq 1$ because the refractive index effect results in incomplete cancellation of the geometrical
effect and the Fitzgerald-Lorentz contraction effect.  Of course it was this cancellation effect that
Fitzgerald and Lorentz actually used to arrive at the length  contraction hypothesis, but they failed to
take the next step and note that the cancellation would be incomplete in the air operated Michelson-Morley
experiment.  In a bizarre development modern Michelson interferometer experiments, which use resonant
cavities rather than interference effects, but for which the analysis here is easily adapted, and with the
same consequences, are operated in vacuum mode.  That denies these experiments the opportunity to see
absolute motion effects.  Nevertheless the experimentalists continue to misinterpret their null results as
evidence against absolute motion.  Of course  these experiments are therefore restricted to merely checking the
Fitzgerald-Lorentz contraction effect, and this is itself of some interest.

All data from  gas-mode interferometer experiments, except for that of Miller,  has been incorrectly 
analysed using only the first effect as in Michelson's initial theoretical treatment, and so the
consequences of the other two effects have been absent.  Repeating the above analysis without these two
effects  we arrive at the  Newtonian-physics time difference which,  for $v << V$, is 
\begin{equation}\label{eqn:QG5}
\Delta t = Ln^3\frac{v^2}{c^3}\cos(2\theta)+\mbox{O}(v^4),
\end{equation}
that is $k^2=n^3$. The value of $\Delta t$, which  is typically of order $10^{-17}s$  in gas-mode
interferometers corresponding to a fractional fringe
shift, is deduced from analysing the fringe shifts, and then    the speed $v_{M}$  has been
extracted  using (\ref{eqn:QG5}), instead of the correct form (\ref{eqn:QG4}) or more generally
(\ref{eqn:QG1}).    However it is very easy to correct for this oversight.  From (\ref{eqn:QG4}) and
(\ref{eqn:QG5}) we obtain for the corrected absolute (projected) speed 
$v_P$ through space, and for $n \approx 1^+$, 
\begin{equation}\label{eqn:QG6}
v_P=\frac{v_{M}}{\sqrt{n^2-1}}.
\end{equation}
For air the correction factor in (\ref{eqn:QG6}) is significant, and even more so for helium.

\vskip12pt
\subsection{The Michelson-Morley Experiment:  1887\label{subsection:themichelsonmorley}}
\vskip6pt

 Michelson and Morley reported  that their interferometer experiment in 1887  gave a
`null-result' which  since then, with rare exceptions, has been claimed to  support the Einstein
assumption that absolute motion has no meaning.  However to the contrary  the Michelson-Morley published
data \cite{MM} shows non-null effects, but much smaller than they expected.  They made observations of
thirty-six  $360^0$ turns  using an $L=11$ meter length interferometer, achieved using multiple reflections,
  operating
in air in Cleveland (Latitude $41^0  30'$N) with six turns near 
$12\!:\!\!00$ hrs ($7\!\!:\!\!00$ hrs ST) on each day of July 8, 9 and 11, 1887  and similarly near
$18\!:\!\!00$ hrs ($13\!\!:\!\!00$ hrs ST) on July 8, 9 and 12, 1887.  Each turn took approximately 6
minutes as the interferometer slowly rotated floating on a tank of mercury.  They published and analysed
the average of each of the  6 data sets.  The fringe shifts were extremely small but within their
observational capabilities.

\begin{figure}[ht]
\hspace{30mm}\includegraphics[scale=0.8]{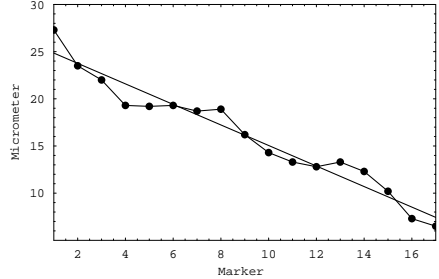}
\caption{{\small   Plot of micrometer  readings for July 11 $12\!\!:\!\!00$ hr ($7\!\!:\!\!00$ ST) showing the
absolute motion induced fringe shifts superimposed on the uniform temperature induced
fringe drift. Most physics books deny that these fringe shifts were seen, despite being
available in \cite{MM}}.\label{fig:MMrawdata}}
\end{figure}

\begin{figure} 
\hspace{12mm}{\small \begin{tabular}{|c|c|c|c|c|c|c|c|c|c|}
\hline
 local &16& 1& 2& 3& 4& 5& 6& 7& \\time & 8  &  9  & 10 &11 &12 & 13 & 14 & 15 & 16 \\
\hline
 12:00hr &27.3 & 23.5 & 22.0 & 19.3& 19.2& 19.3& 18.7& 18.9 &\\July 11 & 16.2 
&14.3& 13.3& 12.8& 13.3& 12.3& 10.2& 7.3& 6.5  \\ 
\hline
 18:00hr &26.0& 26.0& 28.2& 29.2& 31.5& 32.0& 31.3& 31.7& \\July 9 & 33.0
& 35.8& 36.5& 37.3& 38.8&
41.0& 42.7& 43.7& 44.0 \\
\hline
 \end{tabular}}
\vspace{3mm}

{\small Table 2. Examples of Michelson-Morley fringe-shift micrometer readings, from \cite{MM}. 
 The readings for July 11  $12\!\!:\!\!00$ hr are plotted in Fig.\ref{fig:MMrawdata}.}
\end{figure}

\vspace{3mm}

 Table 2 shows   examples of the averaged fringe shift micrometer readings every $22.5^0$ of rotation of
the   Michelson-Morley interferometer \cite{MM} for July 11 12:00 hr  local time and also for July 9
18:00 hr local time.  The orientation of the  stone slab  base is
 indicated by the marks $16,1,2,..$. North is mark 16. The dominant effect was a uniform
fringe drift caused by temporal temperature effects on the length of the arms, and imposed upon that are
the fringe shifts corresponding to the effects of absolute motion, as shown in Fig.\ref{fig:MMrawdata}.

This
temperature effect can be removed by subtracting from the data in each case a best fit to the data of
$a+bk$,
$\{k=0,1,2,.. ,8\}$ for the first $0^0$ to $180^0$  part of each rotation data set.   Then multiplying by
$0.02$ for the micrometer thread calibration   gives the fringe-shift
data points  in Fig.\ref{fig:MMplots}. This factor of $0.02$ converts the micrometer readings to fringe
shifts expressed as fractions of a wavelength.  Similarly a linear fit has been made to the data from the
$180^0$ to $360^0$  part of each rotation data set.  Separating  the full $360^0$ rotation into two $180^0$
parts  reduces the effect of the temperature drift not being perfectly linear in time.

\begin{figure}[ht]
\vspace{20mm}
\hspace{45mm}\setlength{\unitlength}{0.6mm}
\begin{picture}(0,20)
\thicklines
\put(25,10){\vector(-1,0){15}}
\put(25,10){\vector(0,-1){29.5}}
\put(25,10){\vector(-1,2){15}}
\qbezier(0,0)(25,20)(50,0)
\put(30,-15){\Large $\bf v_{in}$}
\put(5,-20){\Large $\bf sun$}
\put(-20,10){\Large ${\bf v}_{tangent}$}
\put(17,30){\Large ${\bf v}_{R}$}
\end{picture}
\vspace{10mm} 
\caption{\small  Orbit of earth about the sun defining the  plane of the ecliptic with tangential orbital
velocity
${\bf v}_{tangent}$ and quantum-foam in-flow velocity  ${\bf v}_{in}$. Then ${\bf v}_{R}={\bf
v}_{tangent}-{\bf v}_{in}$ is the velocity of the earth relative to the quantum foam, after subtracting
${\bf v}_{cosmic}$. }
\label{fig:orbit}
\end{figure}
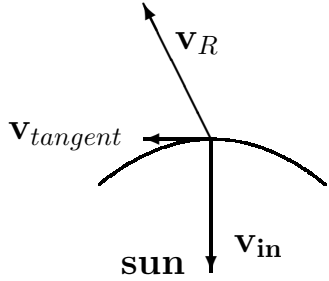

\begin{figure}
\hspace{20mm}\includegraphics[scale=1.0]{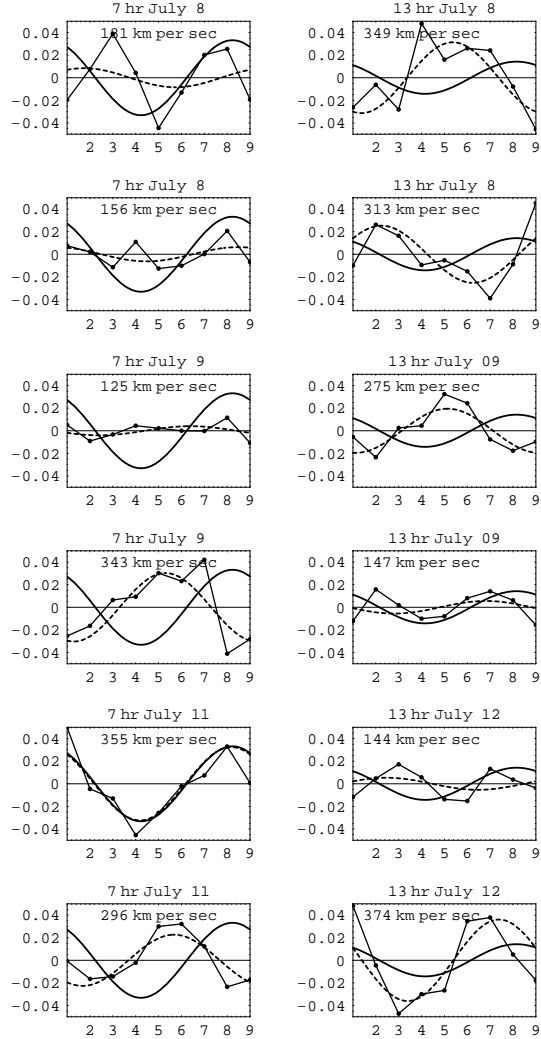}
\caption{{\small Shows all the Michelson-Morley 1887 data after removal of the
temperature induced fringe drifts.  The data for each $360^0$ full turn (the average of 6 individual turns) is
divided into the 1st and 2nd $180^0$ parts and  plotted one above the other.  The dotted curve
shows a best fit to the data, while the full curves show the expected forms using the
Miller direction for ${\bf v}_{cosmic}$.}\label{fig:MMplots}}
\end{figure}

In the new physics there are four main velocities that contribute to the total velocity:
\begin{equation}\label{eqn:QG6b}
{\bf v}= {\bf v}_{cosmic} +{\bf v }_{tangent} -{\bf v}_{in}-{\bf v}_E.
\end{equation}
Here ${\bf v}_{cosmic}$ is the velocity of the solar system relative to some cosmologically defined
galactic quantum-foam system (discussed later) while the other three are local effects: (i) ${\bf v
}_{tangent}$ is the tangential orbital velocity of the earth about the sun,  
(ii) ${\bf v}_{in}$ is  a quantum-gravity radial in-flow  of the quantum foam past the earth towards the
sun, and (iii) the corresponding quantum-foam in-flow into the
earth is ${\bf v}_E$ and makes no contribution to a horizontally operated   interferometer, assuming the
velocity superposition approximation, and also that the
turbulence associated with that flow is not significant. 
The minus signs in (\ref{eqn:QG6b}) arise because, for example, the in-flow towards the sun requires the
earth to have an outward directed velocity against that in-flow in order to maintain a fixed distance from
the sun, as shown in Fig.\ref{fig:orbit}.   For circular orbits  and using in-flow form of Newtonian
gravity the speeds
$v_{tangent}$  and  $v_{in}$ are given by  
\begin{equation}\label{eqn:QG7}
v_{tangent}=\sqrt{\displaystyle{\frac{GM}{R}}},\end{equation}  
\begin{equation}\label{eqn:QG8}
v_{in}=\sqrt{\displaystyle{\frac{2GM}{R}}},\end{equation}
while the net speed $v_R$ of the earth from the vector sum   ${\bf v}_R={\bf v}_{tangent}-{\bf v}_{in}$  is 
\begin{equation}\label{eqn:QG7b}
v_{R}=\sqrt{\displaystyle{\frac{3GM}{R}}},\end{equation}     
where $M$ is the mass of the sun, $R$ is the distance of the earth from the
sun, and $G$ is Newton's gravitational constant. $G$ is essentially a measure of the rate at
which matter effectively `dissipates' the quantum-foam. The gravitational acceleration arises from 
inhomogeneities in the flow.  
 These expressions give $v_{tangent}=30$km/s,  $v_{in}=42.4$km/s and $v_{R}=52$km/s.

 Fig.\ref{fig:MMplots} shows all the data  for the 1887 Michelson-Morley experiment  for the fringe shifts after
removal of the temperature drift effect for each averaged  180 degree rotation. The dotted  curves  come from  
the best fit of $\frac{0.4}{30^2}k^2_{air}v_P^2\cos(2(\theta-\psi))$ to the data. The coefficient
$0.4/30^2$ arises as the apparatus would give a $0.4$ fringe shift, as a fraction of a wavelength,  with
$k=1$ if $v_P=30$ km/s \cite{MM}.  Shown in each figure is the resulting value of $v_P$.  In some cases the
data does not have the expected $\cos(2(\theta-\psi))$ form, and so the corresponding values for $v_P$ are
not meaningful.  The remaining fits give $v_P=331\pm30$ km/s for the   $7\!\!:\!\!00$ hr (ST)
data, and 
$v_P=328\pm50$ km/s for the $13\!\!:\!\!00$ hr (ST) data.   For comparison  the full curves show the
predicted form for the Michelson-Morley data, computed for the latitude of Cleveland, using the Miller
direction (see later) for  ${\bf v}_{cosmic}$ of Right Ascension and Declination $(\alpha=4^{hr} 54^\prime,
\delta=-70^0 30^\prime)$ and incorporating the tangential and in-flow velocity effects for July.   The
magnitude  of the theoretical curves are in general in good agreement with the magnitudes of the
experimental data, excluding those cases where the data does not have the sinusoidal form.  However there are significant
fluctuations in the  azimuth angle.  These fluctuations are also present in the Miller data, and together suggest that
this is a real physical phenomenon, and not solely due to difficulties with the operation of the interferometer.

The Michelson-Morley interferometer data clearly shows the characteristic sinusoidal form with period $180^0$ 
together with a large speed.  Ignoring the effect of the refractive index, namely using the Newtonian value of
$k=1$, gives speeds reduced by the factor $k_{air}$, namely  $k_{air}v_P=0.0241\times 330 $km/s $=7.9$ km/s.
Michelson and Morley reported  speeds in the range $5$km/s - $7.5$km/s. These slightly smaller speeds arise
because they averaged all the  $7\!\!:\!\!00$ hr (ST) data, and separately all the $13\!\!:\!\!00$ hr (ST) data,
whereas here some of the lower quality data have not been used.   Michelson was led to the false conclusion that
because this speed of some $8$ km/s was considerably less than the orbital speed of $30$ km/s  the
interferometer must have failed to have detected absolute motion, and that the data was merely caused by
experimental imperfections.  This was the flawed analysis that led to the incorrect conclusion by Michelson and
Morley that the experiment had failed to detect absolute motion.  The consequences for physics  were extremely
damaging, and are only now being rectified after some 115 years.

\vskip12pt
\subsection{The Miller Interferometer Experiment: 1925-1926\label{subsection:themiller}}
\vskip12pt

Dayton Miller developed and operated  a  Michelson interferometer for over twenty years, with the main sequence of
observations being on Mt.Wilson in the years 1925-1926, with the results reported in 1933 by Miller \cite{Miller2}.  
Accounts of the Miller experiments are available in Swenson \cite{Swenson}.   Miller developed his huge interferometer
over the years, from 1902 to 1906, in collaboration with Morley, and later at Mt.Wilson where the most extensive
interferometer observations were carried out.  Miller was meticulous in perfecting the operation of the interferometer
and performed many control experiments. The biggest problem to be controlled was the effect of temperature changes on
the lengths of the arms. It was essential that the temperature
effects were kept as small as possible, but so long as each turn was performed sufficiently quickly, any
temperature effect could be assumed to have been linear with respect to the angle of rotation. Then a
uniform background fringe drift could be removed,  as in the Michelson-Morley data analysis (see
Fig.\ref{fig:MMrawdata}).  

 In all some 200,000 readings were taken during some 12,000 turns of the
interferometer\footnote{In a remarkable development in 2002 as a result of a visit by James DeMeo to Case
Western Reserve University the original Miller data was located, some 61 years after Miller's death in
1941.  Until then it was thought that the data had been destroyed. Analysis of that data by the author of
this article    has confirmed the accuracy of  Miller's  analysis.  Using more
thorough computer based techniques the data is now being re-analysed.}.  Analysis of the data requires the
extraction of the  speed
$v_M$ and the azimuth angle $\psi$ by effectively fitting the observed time differences, obtained from the
observed fringe shifts,  using (\ref{eqn:QG1}), but with $k=1$.  Miller was of course unaware of the full
theory of the interferometer and so he assumed the Newtonian theory, which neglected both  the
Fitzgerald-Lorentz contraction  and air effects.

\begin{figure}[h]
\hspace{15mm}\includegraphics[scale=1.2]{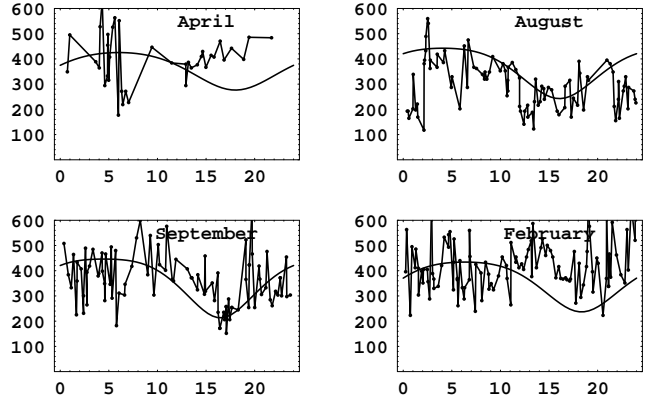}
\caption{ Miller interferometer projected speeds $v_P$ in km/s showing both data and best fit of theory giving $v_{cosmic}=433$ km/s in the
direction ($\alpha=5.2^{hr}, \delta=-67^0$), and using $n=1.000226$ appropriate for the altitude of Mt. Wilson}  
\label{fig:MillerSpeeds}\end{figure}

Miller performed his observations in April, August and September 1925 and February 1926 and the data is  shown in
Figs.\ref{fig:MillerSpeeds}  and  \ref{fig:MillerAz}.   The speeds shown are the Michelson speeds $v_M$, and these
are easily corrected for the two neglected effects by dividing these $v_M$ by $k_{air}=\surd{(n^2-1)}=0.0241$, as in
(\ref{eqn:QG6}). Then for example a speed of $v_M=10km/s$ gives $v_P=v_M/k_{air}=415$km/s.  However this correction
procedure was not available to Miller.  He understood that the theory of the Michelson interferometer was
not complete, and so he introduced the phenomenological parameter $k$ in (\ref{eqn:QG1}). We shall denote
his values by  $\overline{k}$. Miller noted, in fact, that
$\overline{k}^2<<1$, as we would now expect.  Miller then proceeded on the assumption  that ${\bf v}$
should have only two components: (i) a cosmic velocity of the solar system through space, and
(ii) the orbital velocity of the earth about the sun. Over a year this vector sum  would
result in a changing
${\bf v}$, as was in fact observed.  Further, since the  orbital speed was
known, Miller was able to extract from the data the magnitude and direction of ${\bf v}$ as the orbital
speed offered an absolute scale.  For example the dip in the $v_M$ plots for sidereal times $\tau \approx
16^{hr}$ is a clear indication of the direction of ${\bf v}$, as the dip arises at those sidereal
times when the projection $ v_P$ of  ${\bf v}$ onto the plane of the interferometer is at a minimum. During
a 24hr period the value of $v_P$  varies due to the earth's rotation.  As well the $v_M$ plots vary
throughout the year because the vectorial sum of the earth's orbital velocity ${\bf v}_{tangent}$ and the
cosmic velocity  ${\bf v}_{cosmic}$ changes.  There are two effects here as the direction of 
${\bf v}_{tangent}$ is determined by both the yearly progression of the earth in its orbit about the
sun, and also because  the plane of the ecliptic is inclined at $23.5^0$ to the celestial plane. 
Figs.\ref{fig:MillerSpeeds} and  \ref{fig:MillerAz}  show the expected theoretical variation of both
$v_P$ and the azimuth $\psi$  during one sidereal day in the months of April, August, September and
February. These plots show the clear  signature of absolute motion effects as seen in the actual
interferometer data.


Note that the above corrected Miller projected absolute  speed of
approximately  $v_P=415$km/s  is completely consistent with the corrected   projected
absolute speed of some $330$km/s from the Michelson-Morley experiment, though neither Michelson nor Miller were
able to apply this correction.  The difference in magnitude is completely explained by  Cleveland having a
higher latitude than Mt. Wilson, and also by the only two sidereal times of the Michelson-Morley observations. 
So from his 1925-1926 observations Miller had completely confirmed the true validity of the Michelson-Morley
observations and was able to conclude, contrary to their published conclusions, that the 1887 experiment
had in fact detected absolute motion.  But it was too late. By then the  physicists had incorrectly come to
believe that absolute motion was  inconsistent with various `relativistic effects' that had by then been
observed. This was because the Einstein formalism had been `derived'  from the assumption that
absolute motion was  without meaning and so unobservable in principle. Of course the earlier interpretation  of
relativistic effects by Lorentz had by then lost out to the Einstein interpretation.

\begin{figure}[h]
\vspace{5mm}
\hspace{30mm}\includegraphics[scale=0.9]{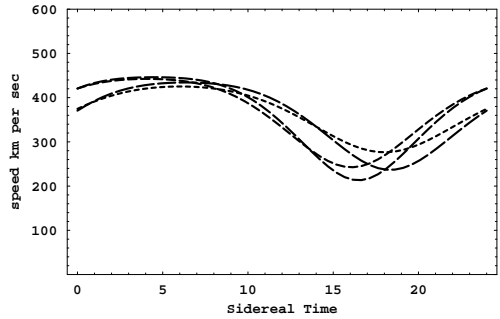}
\caption{\small{ Expected theoretical variation of 
the projected velocity $v_P$ during one sidereal day in the months of April, August, September and
February, labelled by  increasing dash length for  cosmic speed of $433$km/s in the
direction $(\alpha=5.2^{hr}, \delta=-67^0)$}.  This shows the signature of the earth's
orbital rotation.\label{fig:MillerSpTh}}
\end{figure}

\newpage
\begin{figure}[t]
\hspace{15mm}\includegraphics[scale=1.2]{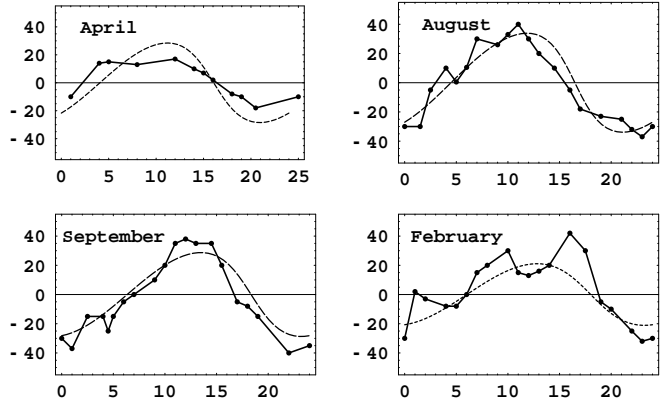}
\caption{ Running time-averaged  Miller azimuths $\psi$, in degrees, measured from south, showing both data and best fit of
theory giving $v_{cosmic}=433$ km/s in the direction ($\alpha=5.2^{hr}, \delta=-67^0$), and using
$n=1.000226$.}  
\label{fig:MillerAz}\end{figure}
\begin{figure}[h]
\vspace{5mm}
\hspace{30mm}\includegraphics[scale=0.9]{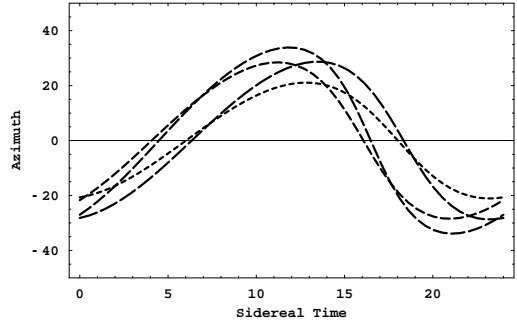}
\caption{\small{ Expected theoretical variation of  the azimuths $\psi$, measured from south, during one sidereal day in the months of April,
August, September and February, labelled by  increasing dash length, for a cosmic speed of $433$km/s in
the direction $(\alpha=5.2^{hr}, \delta=-67^0)$. This shows the signature of the earth's
orbital rotation. }  
\label{fig:MillerAzTh}}\end{figure}

\vskip12pt
\subsection{Gravitational In-flow from the Miller Data\label{subsection:gravitationalinflow}}
\vskip6pt

As already noted Miller was led to the conclusion that for reasons unknown  the existing theory of
the Michelson interferometer did not reveal true values of
$v_P$, and for this reason  he introduced the parameter $k$,  with $\overline{k}$ indicating his numerical
values. Miller had reasoned  that he could determine both ${\bf v}_{cosmic}$ and $\overline{k}$ by observing the
interferometer determined $v_P$ and $\psi$ over a year because the known orbital velocity  of the earth about
the sunwould modulate both of these observables, and by a scaling argument he could determine the absolute
velocity of the solar system.   In this manner he finally determined that $|{\bf v}_{cosmic}|=208$ km/s in the
direction $(\alpha=4^{hr} 54^m, \delta=-70^0 33^\prime)$.  However now that the theory of the Michelson
interferometer has been revealed an anomaly becomes apparent.  Table 3 shows $v=v_M/k_{air}$ 
for each of the four epochs, giving speeds consistent with the revised Michelson-Morley data.  However Table
3 also shows that $\overline{k}$ and the speeds $\overline{v}=v_M/\overline{k}$ determined by the
scaling argument are considerably different.  Here the $v_M$ values arise after taking account of 
the projection effect. That 
$\overline{k}$ is considerably larger than the value of
$k_{air}$ indicates that  another velocity component has been overlooked.   Miller of course only knew of
the tangential  orbital speed of the earth, whereas the new physics predicts that as-well there is a
quantum-gravity radial in-flow ${\bf v}_{in}$ of the quantum foam. We can re-analyse  Miller's data to
extract a first approximation to the speed of this in-flow component.   Clearly it is
$v_R=\sqrt{v_{in}^2+v^2_{tangent}}$ that sets the scale and not
$v_{tangent}$, and because
$\overline{k}=v_M/v_{tangent}$ and $k_{air}=v_M/v_R$ are the scaling relations, then 
\begin{eqnarray}\label{eqn:QG9}
v_{in}&=&v_{tangent}\sqrt{\displaystyle{ \frac{v_R^2}{v_{tangent}^2}-1 }},  \nonumber \\
      &=&v_{tangent}\sqrt{\displaystyle{ \frac{\overline{k}^2}{k_{air}^2}-1 }}.  
\end{eqnarray}

 Using the  $\overline{k}$  values in Table 3 and the value\footnote{In this section we have not modified this
value to take account of the altitude effect  or temperatures atop  Mt.Wilson. This weather information was not
recorded by Miller. The temperature and pressure effect  is that $n=1.0+0.00029\frac{P}{P_0}\frac{T_0}{T}$, where $T$
is the temperature in $^0$K and
$P$ is the pressure in atmospheres.  $T_0=273$K  and
$P_0=$1atm.} of
$k_{air}$  we obtain the
$v_{in}$ speeds  shown in Table 3, which give an average speed of $54$ km/s, compared to the `Newtonian'
in-flow speed of $42$ km/s.  Note that the in-flow interpretation of the anomaly predicts that
$\overline{k}=(v_R/v_{tangent})\,k_{air}=\sqrt{3}\, k_{air}=0.042$. Of course this simple re-scaling of the
Miller results is not completely valid because (i) the direction of ${\bf v}_R$ is of course different to that
of ${\bf v}_{tangent}$, and also not necessarily orthogonal to ${\bf v}_{tangent}$ because of turbulence, and
(ii) also because of turbulence we would  expect some contribution from the in-flow effect of the earth itself,
namely that it is not always perpendicular to the earth's surface, and so would give a contribution to a
horizontally operated  interferometer. 

\begin{figure}[h]
\footnotesize{
\hspace{10mm}\begin{tabular}{|l|l|c|l|l|l|l|} 
\hline 
{\bf Epoch} &\mbox{\ \ }$v_M$ & $\overline{k}$ &$ v=v_M/k_{air}$  & $\overline v=v_M/\overline{k}$ 
&$v=\sqrt{3}\overline{v}$&\mbox{\ \  }
$v_{in}$\\
\hline\hline  
 February   &9.3  & 0.048 & 385.9   & 193.8  & 335.7  &  51.7  \\ \hline
 April   & 10.1 &0.051 &419.1  & 198.0 &342.9  &56.0 \\ \hline 
 August   & 11.2 &0.053 &464.7  & 211.3 &366.0  &58.8 \\ \hline
 September   & 9.6 &0.046 &398.3  & 208.7 &361.5  & 48.8\\ \hline
\hline
\end{tabular}}
\vspace{2mm}

 {\small Table 3. The $\overline{k}$ anomaly,  $\overline{k} \gg k_{air}=0.0241$, as  the
gravitational in-flow effect.
   Here $v_M$ and $\overline{k}$ come from fitting the interferometer data, while $\,v\,$ and
$\,\overline{v}\,$  are
  computed speeds using the indicated scaling. The average of the in-flow speeds
  is $v_{in}=54\pm5$ km/s, compared to the `Newtonian' in-flow speed of $42$ km/s.
 From column 4 we obtain the average $v=417\pm40$km/s.  All speeds in table in  km/s.} 
\end{figure}

An analysis that properly searches for the in-flow velocity
effect clearly requires a complete re-analysis of the Miller data, and this is now possible and underway at
Flinders University as the original data sheets have been found.  It should be noted that the direction
approximately diametrically opposite 
$(\alpha=4^{hr} 54^m, \delta=-70^0 33^\prime)$, namely $(\alpha=17^{hr}, \delta=+68^\prime)$  was at one
stage considered by Miller as being  possible. This is because the Michelson interferometer, being a
2nd-order device, has a directional ambiguity which can only be resolved by using the seasonal motion of the
earth. However as Miller did not include the in-flow velocity effect in his analysis it is possible that a
re-analysis might give this northerly direction as the direction of absolute motion of the solar system.

Hence not only did Miller observe absolute motion, as he claimed,  but the quality and quantity of his data
has also enabled the confirmation of  the existence of the gravitational in-flow effect.  This
is a manifestation of a new theory of gravity and one which relates to quantum gravitational effects via the
unification of matter and space. As well the persistent evidence that this
in-flow is turbulent indicates that this theory of gravity involves self-interaction of space itself.

\begin{figure}
\hspace{25mm}\includegraphics[scale=1.0]{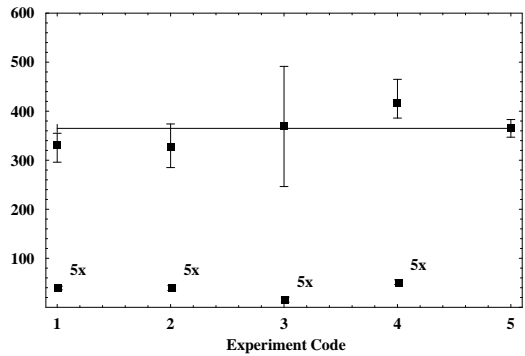}
\caption{\small 
Speeds $v$ in km/s determined from various Michelson interferometer experiments ({\bf 1})-({\bf 4}) and CMB ({\bf
5}): ({\bf 1}) Michelson-Morley (noon observations) and  ({\bf 2})
 ($18^h$ observations) see Sect.\ref{subsection:themichelsonmorley}, ({\bf 3}) Illingworth  
\cite{Illingworth}, ({\bf 4}) Miller, Mt.Wilson
\cite{Miller2}, and finally in ({\bf 5}) the speed from  observations of the CMB
 spectrum dipole term
\cite{CMB}.  The results ({\bf 1})-({\bf 3}) are not
corrected for the
$\pm30$km/s of the orbital motion of the earth about the sunor for the gravitational in-flow speed, though
these corrections were made for ({\bf 4}) with the speeds from Table 3.  The horizontal line at
$v=369$km/s is to aid comparisons with the CMB frame speed data.  The Miller direction is different to the CMB
direction.  Due to the angle between the velocity vector and the plane of interferometer the results ({\bf
1})-({\bf 3})  are less than or equal to the true speed, while the result for ({\bf 4}) is the true speed as
this projection effect was included in the analysis.  These results demonstrate the remarkable consistency
between the three  interferometer experiments. The Miller speed  agrees with the speed from the DeWitte
non-interferometer experiment, in Sect.\ref{subsection:dewitte}. The lower data, magnified by a factor of
5, are the original speeds $v_{M}$  determined from fringe shifts using (\ref{eqn:QG0}) with
$k=1$. This figure updates the corresponding figure in Ref.\cite{CK}.  } 
\label{fig:AllSpeeds}\end{figure}

\vskip12pt
\subsection{The Illingworth Experiment: 1927\label{subsection:theillingworth}}
\vskip6pt

In 1927 Illingworth \cite{Illingworth} performed a Michelson interferometer experiment in which the light
beams passed through the gas  helium,  
\begin{quote}{\it ...as it has such a low index of refraction that variations due to temperature changes are
reduced to a negligible quantity.}
\end{quote}
For helium at STP $n=1.000036$ and so $k^2_{He}=0.00007$, which results in an enormous reduction in sensitivity
of the interferometer.  Nevertheless this experiment gives an excellent opportunity to check the
$n$ dependence in (\ref{eqn:QG6}).   Illingworth, not surprisingly, reported no ``ether drift to an accuracy
of  about one kilometer per second''.  M\'{u}nera \cite{Munera} re-analysed the Illingworth data to obtain a
speed  $v_M=3.13 \pm 1.04$km/s.  The correction  factor in (\ref{eqn:QG6}),
$1/\sqrt{n_{He}^2-1}=118$, is large for helium and gives $v=368\pm 123$km/s. As shown in 
Fig.\ref{fig:AllSpeeds} the Illingworth observations now agree  with those of Michelson-Morley  and  Miller,
though they would certainly be inconsistent without the $n-$dependent correction, as shown in the lower data
points (shown at $5\times$ scale). 

So the use by Illingworth of helium gas, and also by Joos, has turned out have offered a fortuitous opportunity
to confirm the validity of the refractive index effect, though because of the insensitivity of this experiment
the resulting error range is significantly larger than those of the other interferometer  observations.  So
finally it is seen that the Illingworth experiment detected absolute motion with a speed consistent with all
other observations.

\begin{figure}[ht]
\hspace{25mm}\includegraphics[scale=0.9]{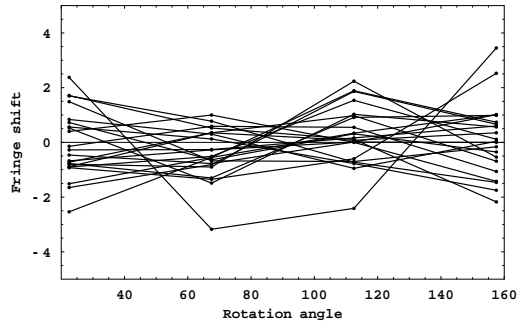}
\caption{ The Joos fringes shifts in $\lambda/1000  $  recorded on  May 30, 1930 from a Michelson interferometer using
helium. Only one of the rotations produced a clean signal of the form expected.} 
\label{fig:JoosData}\end{figure}

\vskip12pt
\subsection{The Joos Experiment: 1930\label{subsection:thejoos}}
\vskip6pt 

\begin{figure}[ht]
\hspace{25mm}\includegraphics[scale=0.9]{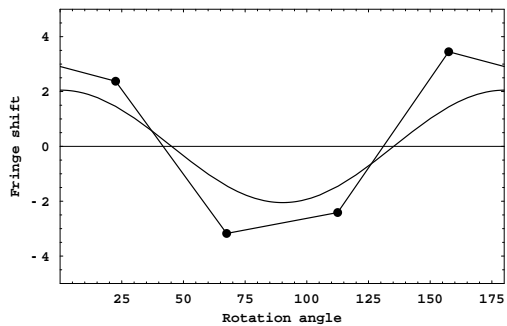}
\caption{  Comparison of the Joos  data for the one  good rotation at $11\mbox{\ } 23^{58}$ with the theoretical prediction
using the  speed and direction from the Miller experiment, together with the length contraction and refractive index
effects. The device sensitivity was $\pm 1$.}  
\label{fig:JoosCompare}\end{figure}

Joos  set out to construct and operate a large vacuum Michelson interferometer at the Zeiss Works in  Jena,
Germany 1930
\cite{Joos}. This interferometer had an effective arm length of 10.5m achieved using multiple refections in
each arm. The vacuum sealing was ineffective and the penetration of air into the vacuum vessel caused
problematic vibrations.  Subsequently Joos used helium, assuming apparently that helium could be
considered as a substitute for a true vacuum\footnote{Thanks to Dr Lance McCarthy for pointing out the
use of helium in this experiment and in extracting the data from the Joos paper.}.  The use of helium is
not mentioned in the Joos paper
\cite{Joos}, but is mentioned by Swenson
\cite{Swenson}.  Joos recorded the fringe shifts photographically, and subsequently analysed the images using a photometer.   The data for 22 
rotations throughout the day of May 30, 1930 are shown in Fig.\ref{fig:JoosData}, and are reproduced from Fig.11 of
\cite{Joos}.  From that data Joos concluded, using an analysis that did not take account of the special 
relativistic length contraction effect, that the fringe shifts corresponded to a speed of only 1.5 km/s.
However as previously noted such an analysis is completely flawed.  As well the data in
Fig.\ref{fig:JoosData} shows that for all but one of the rotations the fringe shifts were  poorly
recorded.  Only in the one rotation, at $11\mbox{\  }23^{58}$,  does the data actually look like the
form expected. This is probably not accidental as the maximum fringe shift was expected at that time,
based on the Miller direction of absolute motion, and the sensitivity of the device was $\pm 1$ thousandth of a
fringe shift. In Fig.\ref{fig:JoosCompare} that one rotation data is compared with the form expected  for Jena on 
May 30 using the Miller speed and direction together with the new refractive index effect,and  using the
refractive index of helium.   The agreement is quite remarkable.  So again contrary the Joos paper and to
subsequent commentators Joos did in fact detect a very large velocity of absolute motion.

\vskip12pt
\subsection{The New Bedford Experiment: 1963\label{subsection:thenewbedford}}
\vskip6pt

\begin{figure}[ht]
\hspace{5mm}\includegraphics[scale=0.8]{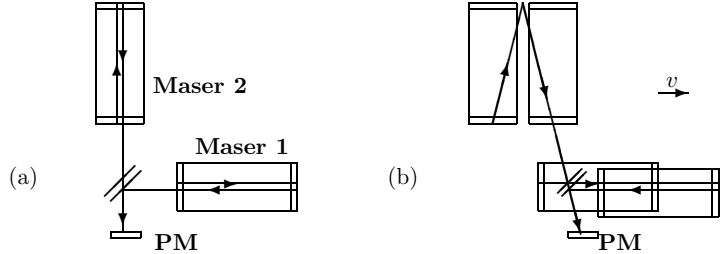}
\vspace{20mm}
\caption{\small{ Schematic diagram for recording the variations in beat frequency between two optical masers:
 (a) when at absolute rest, (b) when in absolute motion at velocity ${\bf v}$. PM is the photomultiplier
detector. The apparatus was rotated back and forth through $90^0$.}  
\label{fig:Masers}}\end{figure}

\begin{figure}[h]
\hspace{25mm}\includegraphics[scale=0.9]{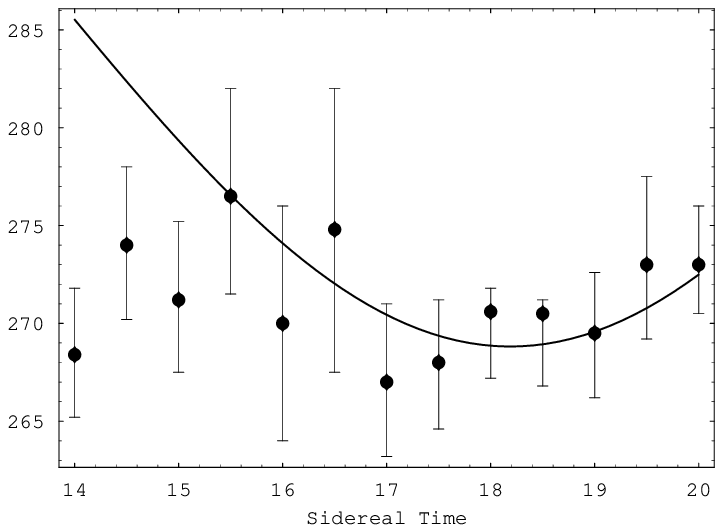}
\caption{\small{  Frequency difference in kHz between the two masers in the 1963 New Bedford experiment after
a $90^0$ rotation. The $275$kHz difference is a systematic repeatable apparatus effect, whereas the superimposed
`dip' at
$17-18\!\!:\!\!00^{hr}$ sidereal time of approximately $3$kHz is a real  time dependent frequency difference. 
The full curve shows the theoretical prediction for the time of the `dip' for this experiment using the Miller
direction for  ${\bf \hat{v}}$ $(\alpha=5.2^{hr}, \delta=-67^0)$ with $|{\bf v}|=433$km/s and
including the earth's orbital velocity  and sun gravitational in-flow velocity effects for January 20, 1963. The
absolute scale of this theoretical prediction was not possible to compute as the refractive index of the He-Ne
gas mixture was unknown. }  
\label{fig:NewBedford}}\end{figure}

In 1964 from an absolute motion detector  experiment at New Bedford, latitude $42^0$N, Jaseja
{\it et al} \cite{Jaseja} reported yet  another `null result'. In this experiment  two He-Ne masers were
mounted with axes perpendicular on a rotating table, see  Fig.\ref{fig:Masers}. Rotation of the table through
$90^0$ produced repeatable variations in the frequency difference of about $275$kHz, an effect  attributed
to  magnetorestriction in the Invar spacers due to the earth's magnetic field.  Observations over some
six consecutive hours on January 20, 1963 from $6\!\!:\!\!00$ am to  $12\!\!:\!\!00$ noon local time  did
produce  a `dip' in the frequency difference of some $3$kHz superimposed on the $275$kHz effect, as shown in
Fig.\ref{fig:NewBedford} in which the local times have been converted to sidereal times.  The most noticeable
feature is that the dip occurs at approximately $17-18\!\!:\!\!00^{hr}$ sidereal time (or $9-10\!\!:\!\!00$ hrs
local time), which agrees with the direction of absolute motion observed by Miller and also by DeWitte (see
Sect.\ref{subsection:dewitte}). It was most fortunate that this particular time period was chosen as at other
times the effect is much smaller, as shown for example 
for the February data in Fig.\ref{fig:MillerSpeeds} which shows the minimum at $18\!\!:\!\!00^{hr}$
sidereal time. The local times were chosen by Jaseja {\it et al} such that if the only motion was due to the
earth's orbital speed the maximum frequency difference, on rotation,    should have occurred at $12\!\!:\!\!00$hr
local time, and the minimum frequency difference at $6\!\!:\!\!00$ hr local time, whereas in fact the minimum
frequency difference occurred at $9\!\!:\!\!00$ hr local time.

As for the Michelson-Morley experiment the analysis of the New Bedford experiment was also bungled. Again this
apparatus can only detect the  effects of absolute motion if the   cancellation between  the
geometrical effects and Fitzgerald-Lorentz length contraction effects is incomplete as  occurs only when the
radiation travels in a gas, here the He-Ne gas present in the maser.  

This double maser apparatus is essentially
equivalent to a Michelson interferometer.  Then the resonant frequency $\nu$ of each maser is
proportional to the reciprocal of the out-and-back travel time. For  maser 1
\begin{equation}\label{eqn:Maser1}
\nu_1=m\frac{V^2-v^2}{2LV\sqrt{1-\displaystyle\frac{v^2}{c^2}}},
\end{equation}
for which a Fitzgerald-Lorentz contraction occurs, while for maser 2
\begin{equation}\label{eqn:Maser2}
\nu_2=m\frac{\sqrt{V^2-v^2}}{2L}.
\end{equation}
Here $m$ refers to the mode number of the masers. When the apparatus is rotated the net observed 
frequency difference is
$\delta
\nu =2(\nu_2-\nu_1)$, where the factor of `2' arises  as the roles of the two masers are reversed after a
$90^0$ rotation. Putting $V=c/n$  we find for $v << V$  and with $\nu_0$  the at-rest resonant frequency,  that
\begin{equation}\label{eqn:Maser3}
 \delta \nu=(n^2-1)\nu_0\frac{v^2}{c^2}+O(\frac{v^4}{c^4}).
\end{equation}
 If we use the Newtonian physics analysis, as in Jaseja {\it et al} \cite{Jaseja}, which neglects both the
Fitzgerald-Lorentz contraction and the refractive index effect, then we obtain  $\delta \nu=\nu_0 v^2/c^2$,
that is without the $n^2-1$ term, just as for the Newtonian analysis of the Michelson interferometer itself.  Of
course the very small magnitude of the absolute motion effect, which was approximately 1/1000 that expected
assuming only  an orbital speed of $v=30$ km/s in the Newtonian analysis,    occurs simply because the
refractive index of the He-Ne gas is very close to one\footnote{It is possible to compare the 
refractive index of  the He-Ne gas mixture in the maser  with the value extractable from this data:
$n^2=1+30^2/(1000\times 400^2)$, or $n=1.0000028$.}. Nevertheless given that it is small  the sidereal time of
the  obvious 'dip' coincides almost exactly with that of the other observations of absolute motion. 
  
 The New Bedford experiment was yet another missed opportunity to have revealed the existence of absolute
motion. Again the spurious argument was that because the  Newtonian physics analysis gave the wrong prediction
then Einstein relativity must be correct.  But the analysis simply failed to take account of  the
Fitzgerald-Lorentz contraction, which had been known since the end of the 19$^{th}$ century,  and  the
refractive index effect which had an even longer history.  As well the authors failed to convert their
local times to sidereal times and compare the time for the `dip' with Miller's time\footnote{There is no
reference to Miller's 1933 paper in  Ref.\cite{Jaseja}. }.

\vskip12pt
\subsection{The DeWitte Experiment: 1991\label{subsection:dewitte}}
\vskip6pt

The Michelson-Morley, Illingworth, Miller, Joos and New Bedford experiments  all used Michelson
interferometers or its equivalent in gas mode, and all revealed  absolute motion.  The Michelson
interferometer is a 2nd-order device meaning that the time difference between the `arms' is proportional to
$(v/c)^2$.  There is also a factor of  $n^2-1$ and for gases like air and particularly
helium  or helium-neon mixes this results in very small time differences and so these experiments
were always very difficult.  Of course without the gas the Michelson interferometer is
incapable of detecting absolute motion\footnote{So why not use a transparent solid in
place of the gas? See Sect.\ref{subsection:solidstate} for the discussion.}, and so there
are fundamental limitations to the use of this interferometer in the study of absolute
motion and related effects. 
 
\begin{figure}
\hspace{30mm}\includegraphics[scale=0.9]{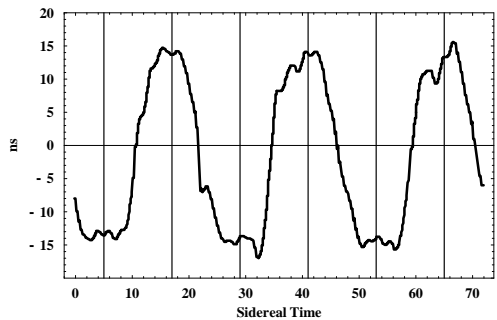}
\caption{\small{ Variations in twice the one-way travel time, in ns, for an RF signal to travel 1.5
km through a coaxial cable between  Rue du Marais and Rue de la Paille, Brussels.  An offset  has been used 
such that the average is zero.  The definition of the sign  convention for $\Delta t$ used by DeWitte
is unclear.  The cable has a
North-South  orientation, and the data is $\pm$ difference of the travel times  for NS and SN
propagation.  The sidereal time for maximum  effect of $\sim\!\!17$hr (or   $\sim\!\!5$hr) (indicated
by vertical lines) agrees with the direction found by Miller and also by  Jaseja {\it et al}, but because of the
ambiguity in the definition of $\Delta t$ the opposite direction would also be consistent  with this data. Plot shows
data over 3 sidereal days  and is plotted against sidereal time. See Fig.\ref{fig:DeWitteTheory}b for
theoretical predictions for one sidereal day. The time of  the year of the data is not identified.
 The fluctuations are evidence of turbulence associated
with  the gravitational in-flow towards the sun. Adapted from DeWitte \cite{DeWitte}.}  
\label{fig:DeWittetimes}}\end{figure}

In a remarkable development  in 1991 a research project  within Belgacom,
the Belgium telecommunications company, stumbled across yet another detection of absolute motion, and one
which turned out to be 1st-order in $v/c$.  The study was undertaken by  Roland DeWitte
\cite{DeWitte}.   This organisation had two sets of atomic clocks in two buildings in Brussels separated by
1.5 km and the research project  was an investigation of  the task of synchronising these two clusters of
atomic clocks. To that end  5MHz radiofrequency signals were sent  in both directions   through two  buried 
coaxial cables linking the two clusters.   The atomic clocks were caesium beam
atomic clocks, and there were three in each cluster. In that way the stability of the clocks could
be established and monitored. One cluster was in a building on Rue du Marais and the second cluster
was due south in a building on Rue de la Paille.  Digital phase comparators were used to measure
changes in times between clocks within the same cluster and also in the propagation times of the RF
signals. Time differences between clocks within the same cluster showed  a linear phase drift caused
by the clocks not having exactly the same frequency together with short term and long term noise.
However the long term drift was very linear and reproducible, and that drift could be allowed for
in analysing time differences in the propagation times between the clusters.

\begin{figure}
\hspace{15mm}
\begin{minipage}[t]{45mm}
\hspace{10mm}\includegraphics[scale=0.6]{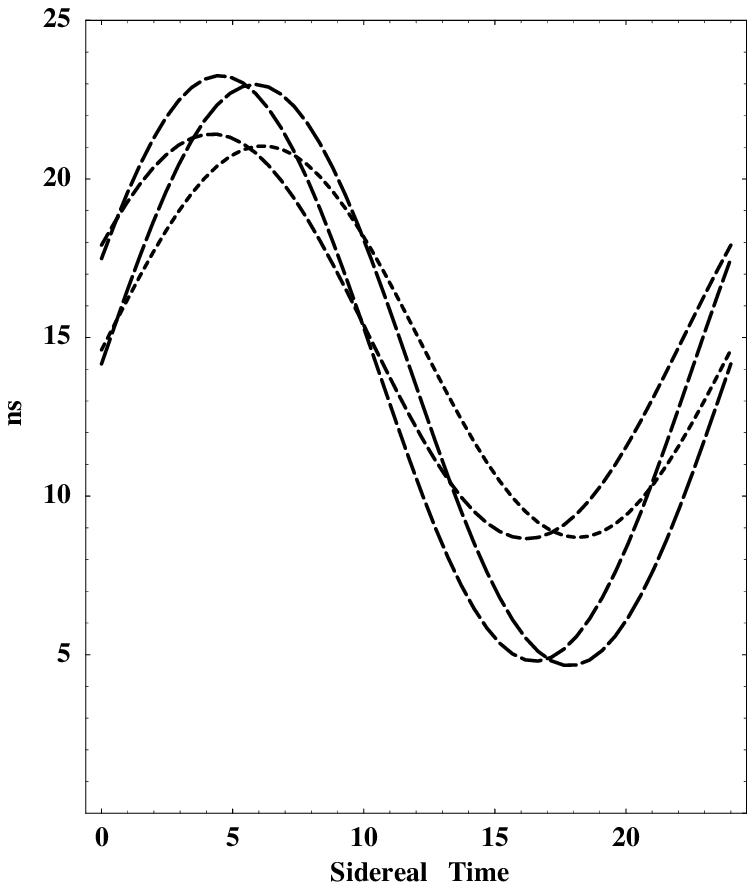}
\makebox[50mm][c]{\small{(a)}}
\end{minipage}
\begin{minipage}[t]{40mm}
\hspace{10mm}\includegraphics[scale=0.6]{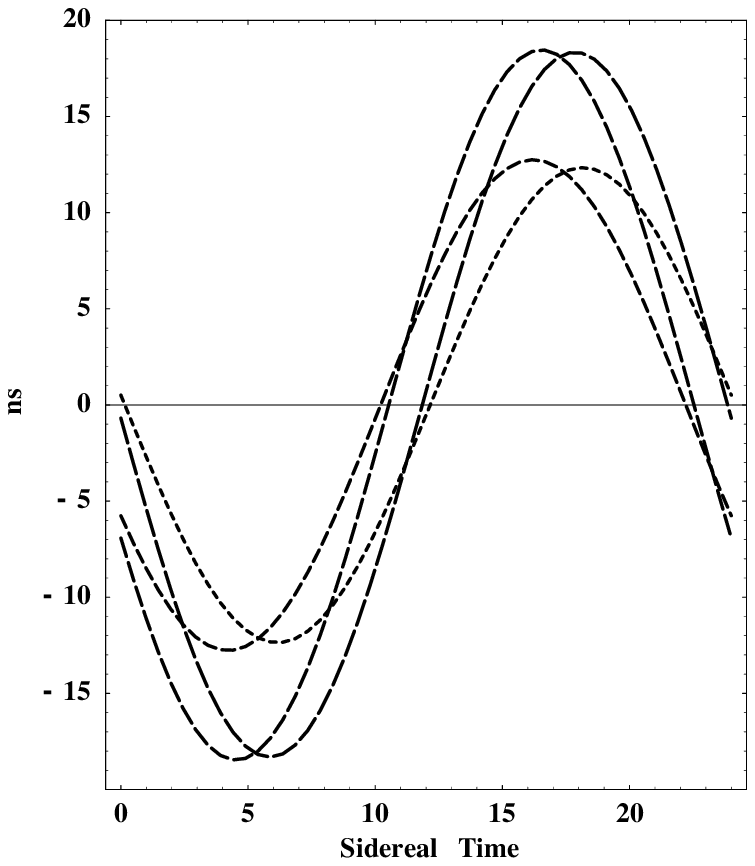}
\makebox[60mm][c]{\small{(b)}}
\end{minipage}
\vspace{0mm}
\caption{\small{ Theoretical predictions for the variations in travel time,  in ns, for
one sidereal day, in the DeWitte Brussels coaxial cable experiment  for ${\bf v}_{cosmic}$ in  the direction
$(\alpha, \delta)=(5.2^h, -67^0)$ and with the Miller magnitude of 443 km/s, and including  orbital and
in-flow effects (but without turbulence). Shown are the results for four days: for the Vernal Equinox,  March
21 (shortest dashes), and for 90, 180 and 270 days later (shown with increasing dash length). 
Figure (a) Shows change in one-way travel time $t_0nv_P/c$ for signal travelling from N to S.  Figure (b)
shows  $\Delta t$, as defined in (\ref{eqn:DW1}), with an offset such that  the average is zero so as to
enable comparison with the  data in Fig.\ref{fig:DeWittetimes}. $\Delta t$ is twice the one-way travel
time. For the direction opposite to
$(\alpha, \delta)=(5.2^h, -67^0)$ the same curves arise except that the identification of the months is
different and  the sign of $\Delta t$ also changes. The sign of $\Delta t$  determines which of the two
directions is the actual direction of absolute motion. However the definition of the sign  convention for
$\Delta t$ used by DeWitte is unclear.}
\label{fig:DeWitteTheory}}\end{figure}

Changes in propagation times  were observed and eventually observations over  178 days were recorded. A sample
of the  data, plotted against sidereal time for just  three days, is shown in Fig.\ref{fig:DeWittetimes}. 
DeWitte recognised that the data was evidence of absolute motion but he was unaware of the Miller experiment 
and did not realise that the Right Ascension for maximum/minimum  propagation time agreed almost
exactly with Miller's direction $(\alpha, \delta)=(5.2^h, -67^0)$. In fact DeWitte expected that the direction
of absolute motion should have been in the CMB direction, but that would have given the data a totally
different sidereal time signature, namely the times for maximum/minimum would have been shifted by 6 hrs. 
The declination of the velocity observed in this DeWitte experiment cannot be determined from the data as
only three days of data are available.   However assuming exactly the same declination as Miller  the speed
observed by DeWitte appears to be also  in excellent agreement with the Miller speed, which in turn is in
agreement with that from the Michelson-Morley and Illingworth experiments, as shown in
Fig.\ref{fig:AllSpeeds}.  

Being  1st-order in $v/c$  the Belgacom experiment is easily analysed to sufficient accuracy by
ignoring relativistic effects, which are 2nd-order in $v/c$.   Let the projection of the absolute
velocity vector ${\bf v}$ onto the direction of the coaxial cable be $v_P$ as before.  Then the
phase comparators reveal the difference  between the propagation
times in NS and SN directions. Consider the analysis with no  Fresnel drag effect,
\begin{eqnarray}
\Delta t &=& \frac{L}{\displaystyle{\frac{c}{n}}-v_P}-
\frac{L}{\displaystyle{\frac{c}{n}}+v_P},\nonumber\\
&=& 2\frac{L}{c/n}n\frac{v_P}{c}+O(\frac{v_P^2}{c^2}) \approx 2t_0n\frac{v_P}{c}.
\label{eqn:DW1}\end{eqnarray}

Here $L=1.5$ km is the length of the coaxial cable, $n=1.5$ is the refractive index of the insulator within
the coaxial cable, so that the speed of the RF signals is approximately $c/n=200,000$km/s, and so
$t_0=nL/c=7.5\times 10^{-6}$ sec is the one-way RF travel time when
$v_P=0$.  Then, for example, a  value of  $v_P=400$km/s would give $\Delta t = 30$ns.  Because Brussels has a
latitude of $51^0$ N then for the Miller direction the projection effect is such that $v_P$ almost varies
from zero to a maximum value of $|{\bf v}|$.  The DeWitte  data in  Fig.\ref{fig:DeWittetimes}
shows $\Delta t$ plotted with a false zero, but  shows a variation of some 28 ns.  So the DeWitte
data is in excellent  agreement with the Miller's data\footnote{There is ambiguity in
Ref.\cite{DeWitte} as to whether the   time variations  in  Fig.\ref{fig:DeWittetimes}  include the
factor of 2 or not, as defined in (\ref{eqn:DW1}). It is assumed here that a factor of 2 is included. }.
The Miller experiment has thus been confirmed by a non-interferometer experiment.

The actual days of the data in 
Fig.\ref{fig:DeWittetimes} are not revealed in Ref.\cite{DeWitte} so a detailed analysis of the DeWitte data
is not possible. Nevertheless theoretical predictions for various days in a year are shown in
Fig.\ref{fig:DeWitteTheory} using the Miller speed of $v_{cosmic}=433$ km/s  and  where the
diurnal effects of the earth's orbital velocity and the gravitational in-flow cause the range of variation of
$\Delta t$ and sidereal  time of maximum effect to vary throughout the year. The predictions give $\Delta t
= 30\pm 4$ ns over a year compared to the  DeWitte value of 28 ns in Fig.\ref{fig:DeWittetimes}.  If
all of DeWitte's 178 days of data were available then a detailed analysis would be possible. 

\begin{figure}[t]
\hspace{30mm}\includegraphics[scale=0.9]{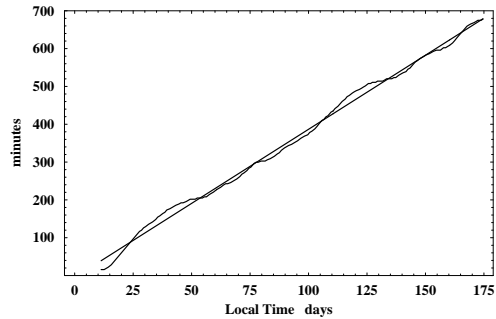}
\caption{\small{ Plot of the negative of the drift of the cross-over time between minimum and
maximum travel-time variation each day (at $\sim10^h\pm1^h$ ST) versus local solar time for some
180 days. The straight line plot is the least squares fit to the experimental data, 
 giving an average slope of 3.92 minutes/day. The time difference between a sidereal day and a solar
day is 3.93 minutes/day.    This demonstrates that the effect is related to sidereal time and not local
solar time.   The actual days of the year are not identified in Ref.\cite{DeWitte}.
 Adapted from DeWitte \cite{DeWitte}.}  
\label{fig:DeWitteST}}\end{figure}

Ref.\cite{DeWitte} does however reveal the sidereal time of the cross-over time, that is a `zero' time
in Fig.\ref{fig:DeWittetimes}, for all 178 days of data.  This is plotted in Fig.\ref{fig:DeWitteST} and
demonstrates that the time variations are correlated with sidereal time and not local solar time.  A
least squares best fit of a linear relation to that data gives that the cross-over time is retarded, on
average, by 3.92 minutes per solar day. This is to be compared with the fact that a sidereal day is 3.93 minutes
shorter than a solar day. So the effect is certainly  cosmological and not associated with any daily thermal
effects, which in any case would be very small as the cable is buried.  Miller had also compared his
data against sidereal time and established the same property, namely that up to  small diurnal effects 
identifiable with the  earth's orbital  motion,  features in the data tracked sidereal time and not
solar time; see Ref.\cite{Miller2} for a detailed analysis.

The DeWitte data is also capable of resolving the question of the absolute direction of motion found by
Miller. Is the direction  $(\alpha, \delta)=(5.2^h, -67^0)$ or the opposite direction? By doing a 2nd-order
Michelson interferometer experiment Miller had to rely on the earth's diurnal effects in order to  resolve
this ambiguity, but his analysis of course did not take account of the gravitational in-flow effect, and so
until a re-analysis of his data his preferred choice of direction must remain to be confirmed.  The DeWitte
experiment could easily resolve this ambiguity by simply noting the sign of $\Delta t$.  Unfortunately it is
unclear  in Ref.\cite{DeWitte}  as to how the sign in Fig.\ref{fig:DeWittetimes} is actually defined, and
DeWitte does not report a direction expecting, as he did, that the direction should have been the same as the
CMB direction.

The DeWitte observations were truly remarkable considering that initially they were serendipitous.  They
demonstrated yet again that the Einstein postulates were in contradiction with experiment.  To my knowledge
no physics journal has published a report of
the  DeWitte experiment. 

That the DeWitte experiment is not a gas-mode Michelson interferometer experiment is very significant.  The
 value of the speed of absolute motion revealed by the  DeWitte experiment of some 400 km/s is in agreement
with the speeds revealed by the new analysis of various  Michelson interferometer data,  which used the
recently discovered refractive index effect, see Fig.\ref{fig:AllSpeeds}. Not only was this effect confirmed
by comparing results for different gases, but the re-scaling of the older $v_M$ speeds to $v=v_M/\sqrt{n^2-1}$
speeds resulting from this effect are now confirmed.

\vskip12pt
\subsection{The Torr-Kolen Experiment: 1981\label{subsection:Thetorr}}
\vskip6pt

\begin{figure}
\vspace{0mm}
\hspace{30mm}\includegraphics[scale=0.8]{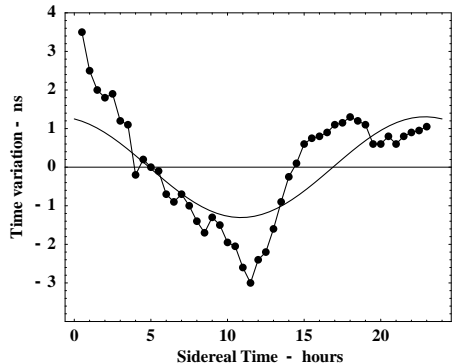}
\caption{\small{Data from the 1981 Torr-Kolen experiment at Logan, Utah \cite{Torr}. 
The data shows variations in travel times
(ns),  for local times,  of an RF signal travelling through 500m of coaxial 
cable  orientated in an E-W direction. Actual days are not indicated but the experiment 
was done during February-June 1981.   Results are for  a typical day.   For the 1st of
February the  local time of $12\!\!:\!\!00$  corresponds to $13\!\!:\!\!00$   sidereal
 time.  The predictions are for   February 
for a cosmic speed of $433$ km/s in the direction  $(\alpha, \delta)=(5.2^h, -72^0)$, and including orbital and in-flow
velocities but without theoretical turbulence.} 
\label{fig:TorrKolen}}
\end{figure} 

A coaxial cable experiment similar to but before the DeWitte experiment was performed at the  University of Utah in 1981 by Torr
and  Kolen \cite{Torr}. This involved two rubidium vapor clocks placed approximately 500m apart with a 5 MHz sinewave
RF signal propagating between the clocks via a nitrogen filled coaxial cable maintained at a constant pressure
of $\sim$2 psi. This means that the Fresnel drag effect is not important in this experiment. Unfortunately the
cable was orientated in  an East-West direction
which is  not a favourable  orientation for observing  absolute motion in the Miller direction, unlike the 
Brussels North-South cable
orientation.  There is no reference to Miller's result in the Torr and  Kolen paper, otherwise they would presumably not have used
this orientation.  Nevertheless  there is a  projection of the absolute motion velocity onto the
East-West cable and Torr and Kolen did observe an effect in that, while the round speed time remained
constant within 0.0001\%c, typical variations in the one-way travel time  were observed,
as shown in Fig.\ref{fig:TorrKolen} by the  data points.  The theoretical  predictions for the 
Torr-Kolen experiment for a cosmic speed of $433$ km/s in the direction  $(\alpha, \delta)=(5.2^h,
-67^0)$, and including orbital and in-flow velocities, are  shown in Fig.\ref{fig:TorrKolen}.  As well the maximum
effect occurred, typically,  at the predicted times.  So the results of this experiment are also in remarkable
agreement with the Miller direction, and the speed of 433 km/s which of course only arises after
re-scaling the Miller speeds for the effects of the gravitational in-flow. As well Torr and Kolen
reported fluctuations in both the magnitude  and time of the maximum variations in travel time just as
DeWitte observed some 10 years later.  Again we argue that these fluctuations are evidence of genuine
turbulence in the in-flow as discussed in Sect.\ref{subsection:inflowturbulence}.  So the Torr-Kolen
experiment again shows strong evidence for the new theory of gravity, and which is over and above its
confirmation of the various observations of absolute motion.

\vskip12pt
\subsection{Galactic In-flow  and the CMB Frame\label{subsection:galacticinflow}}
\vskip6pt
  
Absolute motion (AM) of the solar system has been observed in the direction
$(\alpha,\delta)=(5.2^h, -67^0)$, up to an overall sign to be sorted out,  with a speed
of
$433 $ km/s. This is the velocity after  removing the contribution of the earth's
orbital speed and the sun in-flow effect. It is significant that this velocity is different
to that associated with the Cosmic Microwave Background 
(CMB) relative to which the solar system
has a speed of $369$ km/s in the direction
 $(\alpha,\delta)=(11.20^h,-7.22^0)$, see \cite{CMB}. 
This CMB velocity is obtained by finding the preferred frame in which this thermalised
$3^0$K radiation is isotropic, that is by removing the dipole component.  
The CMB velocity is a measure of the motion
of the solar system relative to the universe as a whole, or at least a shell of the universe
some 15Gyrs away, and indeed the near uniformity of that radiation in all directions
demonstrates that we may  meaningfully refer to the spatial structure of the
universe.  The concept here is that at the time of decoupling of this radiation from
matter that matter was on the whole, apart from small observable fluctuations, at
rest with respect to the quantum-foam system that is space. So the CMB velocity is the
motion of the solar system  with respect to space {\it universally},  but not
necessarily with respect to the   {\it local} space.  Contributions to this  velocity
would arise from the orbital motion of the solar system within the Milky Way galaxy,
which has  a speed of some 250 km/s, and contributions from the motion of the Milky
Way within the local cluster, and so on to perhaps larger clusters.

On the other hand the AM velocity is a vector sum of this {\it universal} CMB
velocity and the net velocity associated with the {\it local} gravitational in-flows into
the Milky Way and the local cluster.  If the CMB velocity had been identical to the AM
velocity then the in-flow  interpretation of gravity would have been proven wrong. We
therefore have three pieces of experimental evidence for this interpretation (i) the
refractive index anomaly discussed previously in connection with the Miller data, (ii) the
turbulence seen in all detections of absolute motion, and now (iii) that the AM velocity is
different in both magnitude and direction from that of the  CMB  velocity, and that this CMB
velocity does not display the turbulence seen in the AM velocity. 

That the AM and CMB velocities are different amounts to the discovery of the resolution 
to the `dark matter' conjecture. Rather than the galactic velocity anomalies being caused by
such undiscovered `dark matter' we see that the in-flow into non spherical galaxies, such as
the spiral Milky Way, will be non Newtonian.   As well it will be interesting to
determine, at least theoretically, the scale of turbulence expected in galactic systems,
particularly as the magnitude of the turbulence seen in the AM velocity is somewhat
larger than might be expected from the sun in-flow alone. Any theory for the turbulence
effect will certainly be checkable within the solar system as the time scale of this
is suitable for detailed observation.

It is also clear that the time of observers  at rest with respect to the CMB frame is 
absolute or  universal time.  This interpretation of 
the CMB frame has of course always been rejected by supporters of the SR/GR formalism.
As for space we note that it has a differential structure, in that different regions
are in relative motion.  This is caused by the gravitational in-flow effect locally, and 
as well by the growth of the universe.

\vskip12pt
\subsection{In-Flow Turbulence and Gravitational 
Waves \label{subsection:inflowturbulence}}
\vskip6pt

\begin{figure}[ht]
\hspace{25mm}\includegraphics[scale=0.9]{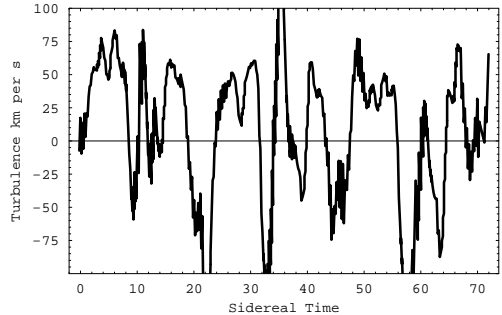}
\caption{\small{ Speed fluctuations  determined from Fig.\ref{fig:DeWittetimes}
by subtracting a least squares best fit of the forms  shown in Fig.\ref{fig:DeWitteTheory}b.
A 1ns variation in travel time corresponds approximately to a speed variation of 27km/s.  The
larger speed fluctuations  actually arise from a fluctuation in the cross-over time, that is, a
fluctuation in the direction of the velocity.  This plot implies that the velocity flow-field
is turbulent.  The scale of this turbulence is comparable to that evident in the Miller data, as
shown in Fig.\ref{fig:MillerSpeeds}  and Fig.\ref{fig:MillerTKTurb}a. }  
\label{fig:Turbulence}}\end{figure}

\begin{figure}
\hspace{5mm}
\begin{minipage}[t]{40mm}
\hspace{10mm}\includegraphics[scale=0.7]{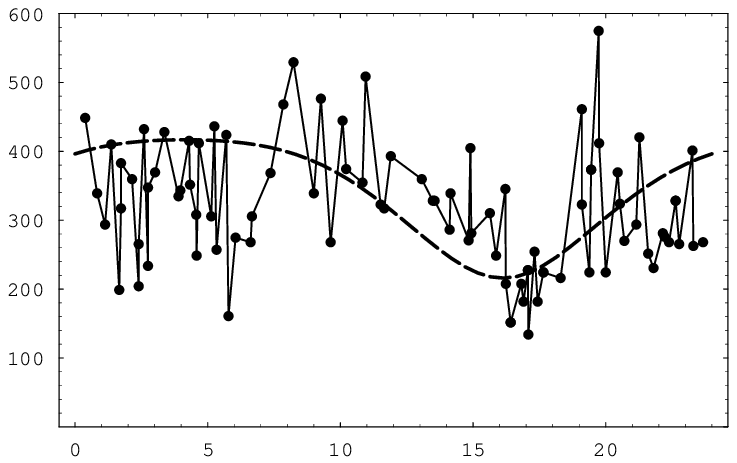}
\makebox[60mm][c]{\small{(a)}}
\end{minipage}
\begin{minipage}[t]{50mm}
\vspace{0mm}\hspace{10mm}\includegraphics[scale=0.7]{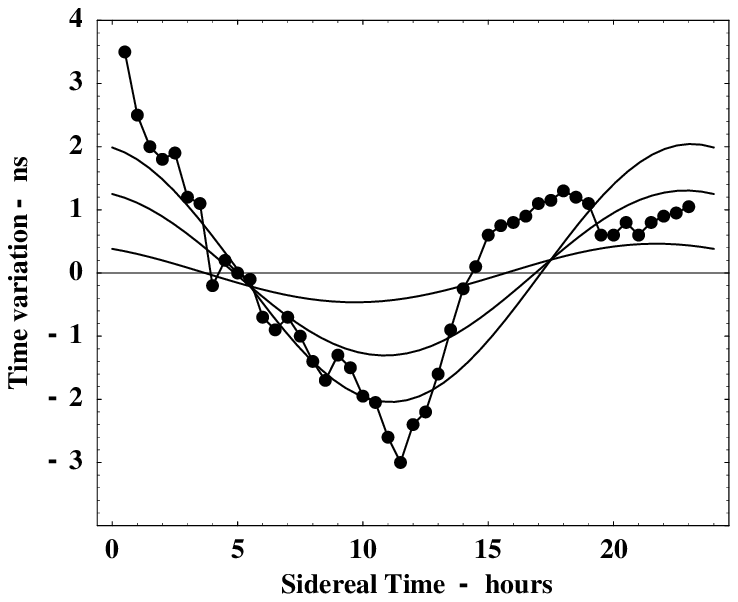}
\makebox[60mm][c]{\small{(b)}}
\end{minipage}
\vspace{0mm}

\caption{\small{ (a) The absolute projected speeds $v_P$ in the Miller experiment  plotted against sidereal time in hours 
for September 1925, showing the variations in speed caused by the gravitational wave turbulence. and (b) 
similar variations in travel times when the declination is varied by
$\pm 10^0$ about the direction $\alpha=5.2^{h}, \delta=-67^0$, for a cosmic speed of 433 km/s in the Torr-Kolen
experiment.}  
\label{fig:MillerTKTurb}}\end{figure}

The velocity  flow-field equation is
expected to have solutions possessing turbulence, that is,  fluctuations in both the
magnitude and direction of the gravitational in-flow component of the velocity flow-field.   
Indeed all the Michelson interferometer experiments showed evidence of such turbulence. The first
clear evidence was from the Miller experiment, as shown in Figs.\ref{fig:MillerSpeeds} and
\ref{fig:MillerAz}.    Miller offered no explanation for these fluctuations  but in his analysis of that data he
did running time averages.  Miller may have in fact have
simply interpreted these fluctuations as purely instrumental effects.  While some of these fluctuations may be
partially caused by weather related temperature and pressure  variations, the bulk of the fluctuations appear to be
larger than expected from that cause alone.   Even the original Michelson-Morley data in
Fig.\ref{fig:MMplots} shows variations in  the velocity field and supports this
interpretation.    However it is significant that the non-interferometer DeWitte data also shows
evidence of turbulence in both the magnitude and  direction of the velocity flow field, as shown
in Fig.\ref{fig:Turbulence}.  Just as the DeWitte data agrees
with the Miller data for speeds and directions the magnitude fluctuations, shown in
Fig.\ref{fig:Turbulence}, are very similar in absolute magnitude to, for example, the Miller speed 
turbulence shown in Fig.\ref{fig:MillerTKTurb}a.  As well the orientation of the Torr-Kolen coaxial cable 
is very sensitive to the directional changes associated with the turbulence. Being almost at
$90^0$ to the direction of absolute motion, any variation in that direction produces significant effects, as shown in Fig. 
\ref{fig:MillerTKTurb}b where the declination is varied by $\pm10^0$.  Indeed Torr and Kolen \cite {Torr} reported significant
fluctuations in the coaxial cable travel times from day to day, as expected.

It therefore  becomes clear that there is
strong evidence from these three experiments for these fluctuations being evidence of  physical turbulence in the flow
field.  The magnitude of this turbulence appears to be somewhat larger than that which would be
caused by the in-flow of quantum foam towards the sun, and indeed following on from
Sect.\ref{subsection:galacticinflow}  some of this turbulence may be associated with galactic
in-flow into the Milky Way.  This in-flow turbulence is a form of gravitational wave and the
ability of gas-mode Michelson interferometers to detect absolute motion means that experimental
evidence of such a wave phenomena has been available for a considerable period of time, but
suppressed along with the detection of absolute motion itself.   Of course flow equations of the
form (\ref{eqn:newgravity}) do not exhibit those gravitational waves of the form that have
been predicted to exist based on the Einstein equations, and which are supposed to propagate at
the speed of light.  All this means that gravitational wave phenomena are very easy to detect and
amount to new physics that can be studied in much detail.

\vskip12pt
\subsection{Vacuum Michelson Interferometers\label{subsection:vacuummichelson}}
\vskip6pt

Over the years vacuum-mode Michelson interferometer experiments have become increasing popular, although the motivation
for such experiments appears to be increasingly unclear. The first vacuum interferometer experiment was planned by Joos
\cite{Joos} in 1930, but because of technical problems helium was actually used, as discussed in section
\ref{subsection:thejoos}. 
 The first actual vacuum experiment was by Kennedy and Thorndike  \cite{KT}.  The result was actually unclear but was consistent with a 
 null effect as  predicted by both the quantum-foam physics and  the Einstein physics.  Only Newtonian physics is
disproved by such experiments.  These   vacuum interferometer experiments do give null results, with increasing confidence level, as for  
example in Refs.\cite{KT,BH,Muller,NewVacuum}, but they only check that the  Lorentz contraction effect completely cancels the geometrical
path-length effect in vacuum experiments, and this is common to both theories. So they are unable to distinguish the new physics from the
Einstein physics.   Nevertheless  recent works
\cite{Muller,NewVacuum} continue to claim that the experiment had been motivated by the desire to look for evidence of absolute
motion, despite effects of 
absolute motion  having been discovered as long ago as 1887. The `null results' are always reported as proof of the
Einstein formalism.   Unfortunately the analysis of the data from such experiments is always by means of the Robertson
\cite{Robertson} and  Mansouri and Sexl formalism \cite{MS}, which purports to be a generalisation of the Lorentz
transformation if there is a preferred frame.  However  we have 
seen that absolute motion effects, that is the existence of a preferred frame, are consistent with the usual Lorentz
transformation, based as it is on the restricted Einstein measurement protocol.  A preferred frame is revealed by
gas-mode Michelson interferometer experiments, and then the refractive index of the gas plays a critical role in
interpreting the data.  The Robertson and  Mansouri-Sexl formalism contains no  contextual aspects such as a refractive
index effect and is thus totally inappropriate to the analysis of so called  `preferred frame' experiments.

It is a curious feature of the history of Michelson interferometer experiments that it went unnoticed that   the
results fell into two distinct classes, namely vacuum and gas-mode,  with recurring  non-null results from
gas-mode interferometers.

\vskip12pt
\subsection{Solid-State Michelson Interferometers\label{subsection:solidstate}}
\vskip6pt

The gas-mode Michelson interferometer has its sensitivity to absolute motion effects greatly reduced by the refractive
index effect, namely the $k^2=n^2-1$ factor in (\ref{eqn:QG0}), and for gases with $n$ only slightly greater than one this
factor has caused much confusion over the last 115 years.  So it would be expected that passing the light beams
through a transparent solid with  $n \approx 1.5$ rather than through a gas would greatly increase the sensitivity. 
Such an Michelson interferometer experiment was performed by Shamir and Fox \cite{ShamirFox} in Haifa in 1969. This
interferometer used light from a He-Ne laser and used perspex rods with  $L=0.26$m. The experiment was
interpreted  in terms of  the   supposed Fresnel drag effect, which has a drag coefficient given by
$b=1-1/n^2$. The light passing through the solid was supposed to be `dragged' along in the direction of 
motion of the solid with a velocity $\Delta {\bf V}=b{\bf v}$ additional to the usual $c/n$ speed. As well  
the Michelson geometrical path difference and the Lorentz contraction effects were incorporated into the
analysis.  The outcome was that no fringe shifts were seen on rotation of the interferometer, and Shamir and
Fox concluded that this negative result {\it``enhances the experimental basis of special relativity''}.

The Shamir-Fox experiment was unknown to us\footnote{This experiment was  performed by Professor Warren Lawrance, an
experimental physical chemist with considerable laser experience.} at Flinders university when in 2002  several meters of
optical fibre were used in a Michelson interferometer experiment which also used a He-Ne laser light source.  Again
because of the $n^2-1$ factor, and even ignoring the Fresnel drag effect, one would have expected large fringe shifts on
rotation of the interferometer, but none were observed.  As well in a repeat of the experiment single-mode optical fibres
were also used and again with no rotation effect seen.   So this experiment is consistent with the Shamir-Fox experiment. 
Re-doing the analysis by including the supposed   Fresnel drag effect, as Shamir and Fox did, makes no material
difference to the expected outcome.  In combination with the non-null results from the gas-mode interferometer
experiments along with the non-interferometer experiment of DeWitte it is clear that transparent solids behave differently
to a gas when undergoing absolute motion through the quantum foam.  Indeed this  in itself is a discovery of a new
phenomena.   

The most likely explanation is that the physical Fitzgerald-Lorentz contraction effect has a anisotropic effect on the
refractive index of the transparent solid, and this is such as to cause a cancellation of any differences in travel time
between the two arms  on rotation of the interferometer.  In this sense a transparent solid medium shares this
outcome with the vacuum itself. 
    
\vskip12pt
\subsection{Absolute Motion and Quantum Gravity\label{subsection:absolutmeqg}}
\vskip6pt
 
Absolute rotational motion had been recognised as a meaningful and observable phenomenon from the very beginning of
physics. Newton had used his rotating bucket experiment to illustrate the reality of absolute rotational motion,
and later Foucault  and Sagnac provided further experimental proof. But for absolute linear motion the history
would turn out to be completely different. It was generally thought that absolute linear motion was undetectable,
at least until Maxwell's electromagnetic theory appeared to require it. In perhaps the most bizarre sequence of
events in modern science it turns out that absolute linear motion has been apparent within experimental data for
over 100 years. It was missed in the first experiment designed to detect it and from then on for a variety of
sociological reasons it became a concept rejected  by physicists and banned from their journals despite
continuing  new experimental evidence. Those who pursued the scientific evidence were treated with scorn and
ridicule.  Even worse was  the impasse that this obstruction of the scientific process resulted in, namely the
halting of nearly all progress in furthering our understanding of the phenomena of gravity. For it is clear from
all the experiments that were capable of detecting absolute motion that there is present in that data evidence of
turbulence within the velocity field.  Both the in-flow itself and the  turbulence are manifestations at a
classical level of what is essentially  quantum gravity processes, for these processes imply
that space has structure. 

Process Physics has given a unification of explanation and description of physical phenomena based upon the limitations of
formal syntactical systems which had nevertheless achieved a remarkable encapsulation of many phenomena, albeit in a
disjointed and confused manner, and with a dysfunctional ontology  attached for good measure.  As argued in
\cite{RC01} space is a quantum system continually classicalised by on-going non-local collapse
processes.  The emergent phenomenon is foundational to existence and experientialism.   Gravity in this system is
caused by  differences in the
 rate of processing of the cellular information within the network which we experience as space, and
consequentially there is a differential flow of information which can be affected by the presence of matter or even by
space itself.  Of course the motion of matter including photons with respect to that spatial information content  is
detectable because it affects the geometrical and chronological attributes of that matter, and the experimental evidence
for this has been exhaustively discussed  above.   What has become very clear is that the
phenomenon of gravity is only understandable once we have this unification of the quantum phenomena of matter and
the quantum phenomenon of space itself. In Process Physics the difference between matter and space is subtle. It
comes down to the difference between informational patterns that are topologically preserved and those
information patterns that are not.    One outcome of this unification is that as a consequence of having a
quantum phenomenon of space itself we obtain an informational explanation for gravity, and which at a suitable
level  has an emergent quantum description.  In this sense we have an emergent quantum theory of gravity.  Of
course no such quantum description of gravity  is derivable from quantising Einsteinian gravity itself. This
follows on two counts, one is that the Einstein gravity formalism  fails on several levels, as discussed
previously, and second that quantisation has no validity as a means of uncovering deeper physics.   Most
surprising of all is that   having uncovered the logical necessity for gravitational phenomena it also appears
that even the seemingly well-founded Newtonian account of gravity has major failings.  The denial of this
possibility has resulted in an unproductive search for dark matter.  Indeed like dark matter and spacetime much
of present day physics has all the hallmarks of another episode of Ptolemy's epicycles, namely concepts that
appear to be well founded but in the end turn out to be illusions, and ones that have acquired the status of
dogma.

If the Michelson-Morley experiment had been properly analysed and the phenomena revealed by the data exposed, and this
would have required in 1887 that Newtonian physics be altered, then as well as the subsequent path of physics being very
different, physicists would almost certainly have discovered both the gravitational in-flow effect and associated  
 gravitational waves.  

It is clear then that observation and measurement of absolute motion leads directly to a
changed paradigm regarding the nature and manifestations of gravitational phenomena.   There are two aspects
of such an experimental program. One is the characterisation of the turbulence and its linking to the new
non-linear term in the velocity field theory. This is a top down program. The second aspect is a bottom-up
approach where the form of the velocity field theory, or its modification, is derived from the deeper
informational process physics.  This is essentially the quantum gravity route.   The turbulence is of course
essentially a gravitational wave phenomena and networks of 1st-order interferometers will permit  spatial
and time series analysis.  There are a number of other  gravitational anomalies which may also  now be
studied using such an interferometer network, and so much new physics can be expected to be uncovered.

\vskip12pt
\section{ Conclusions\label{section:conclusions}}
\vskip6pt

We have shown here that seven experiments, so far, have clearly revealed experimental evidence of absolute motion. As
well these are all consistent with respect to the direction and speed of that motion.  This clearly refutes the
fundamental postulates of the Einstein reinterpretation of the relativistic effects that had been developed earlier
by Lorentz and others.  Indeed these experiments are consistent with the Lorentzian interpretation of the special
relativistic effects in which reality displays both absolute motion  and relativistic effects. It is absolute
motion that actually causes these relativistic effects. Data from the five Michelson interferometer fringe-shift
experiments had never been properly analysed until now. That analysis requires that the Fitzgerald-Lorentz
contraction effect be taken into account, as well as the effect of the gas on the speed of light in the
interferometer.  Only then does the fringe-shift data from air and helium interferometer experiments become
consistent, and then also consistent with the two RF coaxial cable travel-time  experiments.   The seasonal changes
in the Miller fringe-shift data reveal the orbital motion of the earth about the sun, as well as an in-flow of space
past the earth into the sun. These results
 support the new theory of gravity. As well the large cosmic velocity of the solar system is seen to be different to
the velocity associated with the Cosmic Microwave Background, which implies another gravitational in-flow, this
time into the Milky Way and Local Cluster.  The fringe-shift data  has also indicated the presence of turbulence in these
gravitational in-flows, and this amounts to the detection of gravitational waves. These are waves predicted by the new
theory of gravity, and not those associated with the Hilbert-Einstein theory of gravity.  As noted  the
Newtonian theory of gravity is deeply flawed, as revealed by its inability to explain a growing number of
gravitational anomalies, but which are explained by the new theory. These flaws arose because the solar system
was too special, because of its high spherical symmetry,   to have revealed the full range of phenomena that
is gravity. General Relativity `inherited' these flaws, and so is itself flawed.  As discussed 
the clear-cut checks of General Relativity were actually done in systems also with high spherical symmetry.

 Here  a new theory of gravity has been proposed.  It passes all the key existing tests, including the operation
of the GPS,  and also appears to be capable of explaining numerous gravitational anomalies.   The phenomena
present in these anomalies provide opportunities for further tests of the new gravitational physics, as
illustrated here by the mine/borehole $g$ anomaly.  This new theory explains why elliptical galaxies display
a very small `dark matter' effect, in comparison with the large effect  for the spiral galaxies.  This new
theory is supported by the Miller, Torr and Kolen, and DeWitte  absolute motion experiments in that they reveal
the turbulent in-flow of space associated with gravity, namely the discovery of those gravitational waves
predicted by the new theory,  as well as the existence of absolute motion itself.  

 The experiments are
consistent with the Lorentzian interpretation of the special relativistic effects in that reality displays both
absolute motion effects {\it and} relativistic effects.  It is absolute motion that causes the special relativistic
effects.     We saw
that the Galilean transformation together with the  absolute motion effects of time dilations and length contractions
for moving matter systems leads to  the data from  observers in absolute motion being related by the Lorentz
transformation, so long as their data are not corrected for the effects of absolute motion.  So the new Process Physics
brings together transformations that were, in the past, regarded as mutually exclusive.  Essentially the quantum foam
system that is space generates phenomena that are more subtle than currently considered in physics. Discussed in
\cite{URL,RC01} was an explanation for the `dark energy' phenomena, previously known as the cosmological constant,
associated with the accelerating rate of expansion of the universe - that also is seen to be a consequence of the
information theoretic aspects of the quantum foam.

\vskip12pt
 \section{ Acknowledgments}
\vskip6pt

Special thanks to Professor Warren Lawrance, Professor Igor Bray, Dr Ben Varcoe and Dr Lance McCarthy.
Thanks to Katie Pilypas for running the codes to fit the Miller data.

\end{document}